# Thin-Film Stabilization and Magnetism of $\eta$-Carbide Type Iron Nitrides

Baptiste Julien[1]*, Abrar Rauf[2], Liam A. V. Nagle-Cocco[3], Rebecca W. Smaha[1], Wenhao Sun[2], Andriy Zakutayev[1], Sage R. Bauers[1]*

1. *Materials Science Center, National Laboratory of the Rockies, Golden, CO 80401, United States.*
2. *Department of Materials Science and Engineering, University of Michigan, Ann Arbor, MI 48109-1079, United States.*
3. *Stanford Synchrotron Radiation Lightsource, Stanford University, Menlo Park, CA 94025, United States*

## Abstract

Transition-metal nitrides in $\eta$-carbide type structures exhibit unusual bonding motifs and proximity to magnetic instabilities. Yet they remain unexplored in thin-film form due to the difficulty of stabilizing nitrogen-poor ternaries among competing phases. Here, we report the thin-film synthesis and phase-stability mapping of the $\eta$-nitride systems Fe-W-N and Fe-Mo-N. Amorphous Fe-M-N (M = W, Mo) combinatorial libraries deposited by reactive co-sputtering crystallize upon rapid thermal annealing, enabling systematic identification of synthesis windows as a function of composition and annealing temperature. Using laboratory powder X-ray diffraction and synchrotron grazing incidence wide angle X-ray scattering, we establish that $Fe_3Mo_3N$-based $\eta$-carbide phases form over a substantially broader compositional and thermal range than W-based compositions, where $\eta$ structures are stabilized only when the films are Fe-rich. These trends are rationalized using mixed chemical-potential vs. composition phase diagrams that capture the narrow nitrogen chemical-potential stability of $\eta$-nitrides. Magnetic measurements reveal that ferromagnetism is induced in Fe-rich $Fe_{3.54}Mo_{2.46}N$ with a small exchange-bias-like response that is absent in $Fe_3W_3N$-based compositions, highlighting the sensitivity of magnetic behavior to modest deviations from stoichiometry. This work establishes practical thin-film synthesis routes for $\eta$-nitride materials and demonstrates how composition can be tuned to access emergent magnetic phenomena in these complex nitrides.

## Introduction

Ternary transition metal nitrides (TMNs) with nitrogen-poor stoichiometries (i.e., involving non $d^0$ TMs) represent an interesting class of materials due to their combination of strong metallic bonding and covalent interactions, giving rise to diverse functional properties ranging from exceptional hardness and thermal stability [1,2] to superconductivity [3], magnetism [4], and catalytic activity [5]. In thin films, we previously showed that nitrogen-rich TMNs can be stabilized under high nitrogen chemical potential ($\mu_N$), such as plasma-cracked $N_2$, where strongly nitriding conditions enable phases that are metastable in bulk form [6–9]. We leveraged this strategy to predict and experimentally realize a broad class of new stable and metastable nitrogen-rich transition metal nitride semiconductors [8,10–12]. In contrast, stabilizing nitrogen-poor TMN thin films presents a fundamentally more challenging problem, as it requires precise control of intermediate $\mu_N$ in the presence of competing nitride and intermetallic phases. This is amplified in multielement systems, where the number of thermodynamically accessible competing phases grows rapidly. Despite this narrow $\mu_N$ window, many nitrogen-poor nitrides are thermodynamically stable owing to unique bonding mechanisms, including reductive stabilization, that favor low-nitrogen compositions in multielement system [8]. These effects open chemically rich regions of nitride phase space but realizing them experimentally requires precise control and understanding of synthesis pathways, particularly in thin-film form.

$\eta$-carbide-type TMNs (referred to as '$\eta$-nitrides' in this work) such as $Fe_3W_3N$ and $Fe_3Mo_3N$ occupy a particularly interesting region of this design space. These compounds are structurally analogous to the well-known $\eta$-carbides $Fe_3W_3C$ and $Fe_3Mo_3C$, with nitrogen occupying the interstitial sites of the metal framework in place of carbon. This substitution modifies the bonding environment and electronic structure, which directly impacts the mechanical, electronic and magnetic properties [13]. While the carbide analogues have been extensively studied as hard coating [14,15], the corresponding nitrides remain relatively underexplored. Importantly, despite extensive bulk investigations, $\eta$-nitride thin-films have not been reported or systematically studied.

Previous bulk studies on Fe-containing $\eta$-nitrides have revealed complex magnetic behavior that is highly sensitive to slight changes in composition [13,16–22]. Notably, $Fe_3Mo_3N$ was found to be located near a quantum critical point, where ferromagnetic (FM) order can be induced from the non-magnetic ground state through minimal Co substitution [18]. These observations position $\eta$-nitride materials as a compelling platform for exploring emergent magnetism driven by subtle changes in composition. However, the lack of thin-film synthesis routes and combinatorial approaches has not facilitated the systematic exploration of composition-structure-property relationships in these materials, as metastable or off-stoichiometric regimes are often more easily accessible in thin films.

Here, we employ a combinatorial thin-film approach to synthesize amorphous, compositionally-graded Fe-W-N and Fe-Mo-N samples via reactive magnetron co-sputtering at room temperature, followed by post-deposition rapid thermal annealing to induce crystallization. Using laboratory X-ray diffraction (XRD) and synchrotron grazing-incidence wide-angle scattering (GIWAXS), we map the phase evolution of both systems as a function of composition and annealing temperature. These experimental results are interpreted using calculated chemical-potential vs. composition phase diagrams that capture the narrow stability windows of these $\eta$- nitrides. We identify that $Fe_3Mo_3N$ has a wider synthesis window than $Fe_3W_3N$. Moreover, we show that minor stoichiometry deviations yield pronounced changes in magnetic behavior, including inducing ferromagnetism and exchange-bias effects in Fe-rich $Fe_{3.54}Mo_{2.46}N$. Together, these results establish practical thin-film synthesis routes for $\eta$-nitrides and demonstrate the sensitivity of their magnetic behavior to modest off-stoichiometry.

## Materials and Methods

### Synthesis

Thin film amorphous Fe-W-N and Fe-Mo-N precursors were deposited using radio-frequency (RF) reactive magnetron co-sputtering. High-purity elemental 2” targets of Fe and W or Mo (Kurt J. Lesker, 99.95% purity) were co-sputtered in an $Ar/N_2$ atmosphere where the Ar and $N_2$ flows were set to 8 and 1 sccm, respectively. The RF power of all the source targets was set to 40 W. The base pressure was maintained below $3 \times 10^{-7}$ Torr, and the pressure during deposition was set to 8 mTorr. Films were deposited on 2” square Si substrates (50.8 × 50.8 mm) coated with a $SiN_x$ layer of ~100 nm to prevent any chemical reaction with Si during annealing experiments. The two magnetrons were positioned in a confocal geometry, and the substrate was maintained stationary to create a composition gradient across the film in one direction (see Figure 1a). The depositions were performed without external heating, considering the self-heating effect from the plasma. After each deposition, the 2” square combinatorial libraries were first characterized by X-ray fluorescence (XRF) and then cut into stripes along the gradient direction, for annealing experiments. Post-deposition annealing was performed in a mini lamp annealer (Advance Riko, MILA-5000) in flowing nitrogen atmosphere with a flow of 10 slpm. The samples were first heated to 100 °C for 3 min under $N_2$ to evaporate surface contaminants and absorbed water and then rapidly ramped up to the desired temperature with a heating rate of 20-30 °C/s. The temperature was maintained for 20 min. At the end, the films were allowed to cool naturally in $N_2$ atmosphere.

**Characterization**

The metal-to-metal ratio was characterized by XRF on a Bruker M4 Tornado using an Rh X-ray source operating at 50 kV and 200 μA. XRF spectra were acquired with an exposure time of 120 s and a spot size of 25 μm. An automated stage was used record spectra across the 2” square libraries, with a horizontal step of 4 mm and a vertical step of 12.5 mm. The metal atomic ratio in the $\eta$-nitrides is defined as $x$ = Fe/(Fe+M), where Fe and M represent the atomic fractions of Fe and W or Mo respectively. Crystalline phases across the films were first screened by laboratory X-ray diffraction (XRD), a Bruker D8 diffractometer equipped with an HI-STAR area detector and operating with a Cu Kα source (λ = 1.54 Å). A mapping stage was used to sequentially acquire diffraction 2D images across the combinatorial samples, with an exposure of 225 s for each point and a 4 mm separation between each point. Samples of interest were further characterized by grazing-incidence wide-angle X-ray scattering (GIWAXS) at the Stanford Synchrotron Radiation Lightsource (SSRL, beamline 11-3), using a 12.7 keV radiation source (λ = 0.97625 Å) and a Rayonix MX225 CCD area detector. The data were collected with a 3° incident angle, a sample-to-detector distance of 150 mm, and a spot size of 50 μm by 150 μm. Structural refinements by a LeBail method were performed with GSAS-II [23].

Magnetic properties were measured in a Quantum Design DynaCool Physical Property Measurement System (PPMS). ~5×5 mm samples of interest were cut from the combinatorial stripes and mounted on quartz paddles using GE varnish. DC magnetization was measured upon warming from 2 K to 300 K under a field of 5 kOe (0.5 T) applied perpendicular to the sample surface, in both field-cooling (FC) and zero field-cooling (ZFC) modes. Field-dependent magnetization loops were recorded from -10 kOe to 10 kOe at different temperatures. To isolate the contribution of the film from the substrate, a piece of bare substrate was measured in the same configuration and then subtracted to the sample signal, applying a scale factor to account for mass difference. More details about the processing of magnetic data employed in this work can be found in the Supplementary Material.

**Calculations**

We first assembled a comprehensive list of Fe-W-N and Fe-Mo-N phases from the Materials Project (MP), using the MP API [24] to filter out entries that were hypothetically generated by crystal structure prediction but have not been experimentally observed such as the $W_2N_3$ phase (mp-1216242) which along with similar phases over-stabilize the ternary hull thereby removing experimentally confirmed phases like hexagonal WN (mp-991) [6,25]. For each remaining compound, Density Functional Theory (DFT) calculations were carried out with the Vienna Ab Initio Software Package [26] (VASP) using Pymatgen's [27] MPRelaxSet to ensure consistency across calculations. The resulting ground state energies were subsequently refined through the Materials Project correction scheme [28,29] to obtain more accurate formation energies for assessing relative phase stability. These energies were then used to construct mixed nitrogen chemical potential – metallic composition phase diagrams by using the Legendre transformed intensive potential given by the following equation,

$$\phi(\mu_{Fe}, \mu_M, \mu_N) = E(N_{Fe}, N_M, N_N) - \mu_{Fe}N_{Fe} - \mu_M N_M - \mu_N N_N \quad (1)$$

where M = Mo/W, $\mu_x$ represents the chemical potential of substance $x$, and $N_x$ is the number of $x$ atoms. The lower half-space intersection of this energy hyper-surface in $\phi - \mu_{M_1} - \mu_{M_2} - \mu_N$ space is used to obtain the stability domains of each phase. The nitrogen chemical potential stability window is then obtained by projecting this stability domain on to the $\mu_N$ axis. Finally, the 2-phase and 3-phase coexistence regions are then extracted by identifying domains with shared vertices. Further details of the implementation can be found in the Supplementary Materials.

## Results and Discussion

### Accessing the $\eta$-Nitride Phase through Thermal Annealing of Amorphous Nitride Precursors

Combinatorial deposition of Fe-*M*-N (*M* = W, Mo) on Si/$SiN_x$ substrates without external heating yielded films with the 1:1 Fe:*M* composition near the center of the compositionally graded substrates, as measured with XRF (Figure S1). Figure 1b and 1c present XRD patterns collected along the composition gradient of the combinatorial films. The absence of diffraction peaks in both cases indicates that the films are amorphous across the composition range investigated and no crystalline phases are formed. The broad "humps" observed around $q$ = 2.8 $Å^{-1}$ probably correspond to some short-range ordering from metal-nitrogen bonds. As shown in Figure S3a, the as-grown films show metallicity with electrical resistivity around 100 – 350 μΩ·cm, which is in the reported range for Fe-based amorphous alloys [30]. The resistivity only slightly increases with Fe content in amorphous Fe-W-N films, whereas it significantly decreases in Fe-Mo-N.

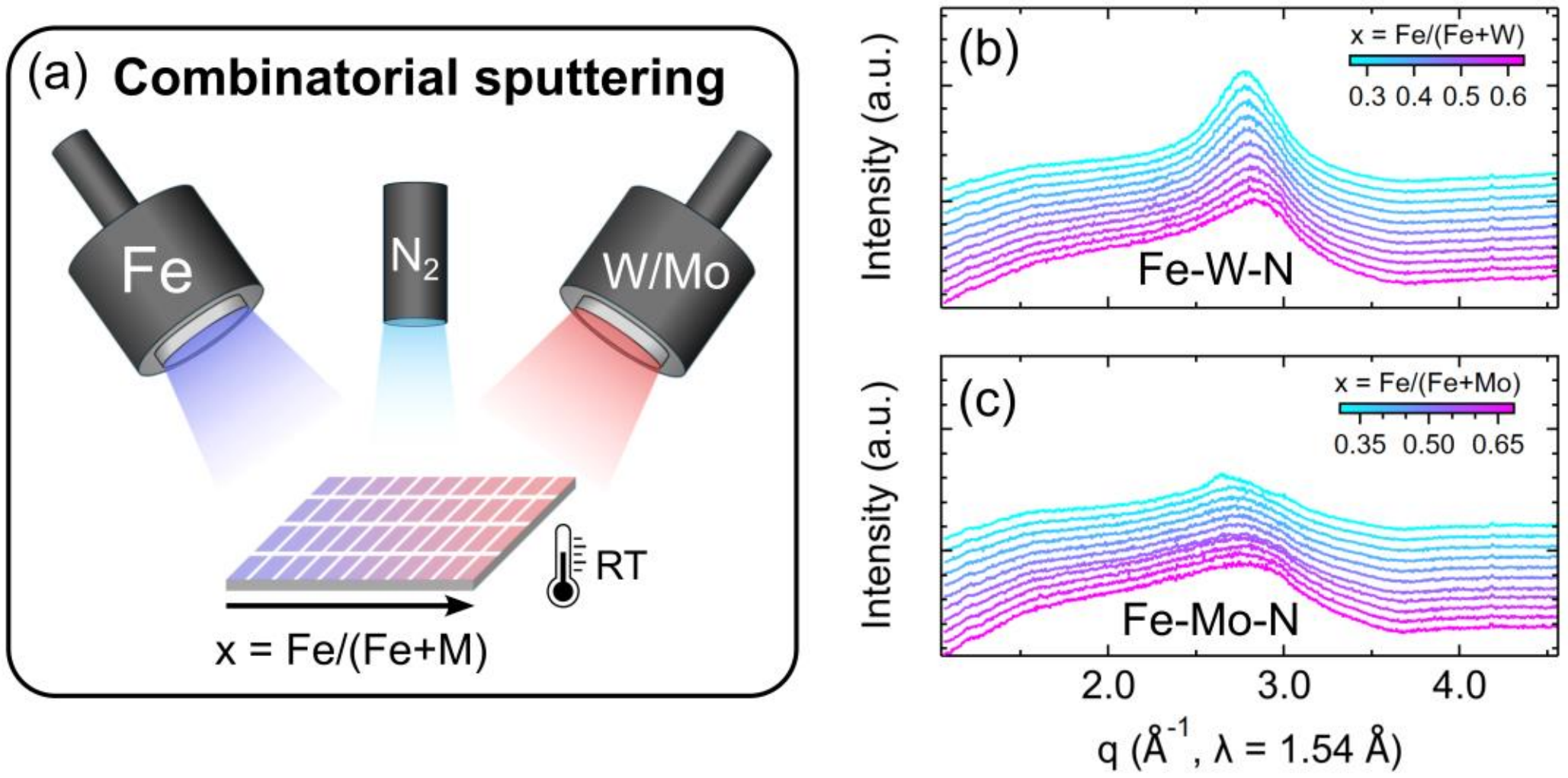

**Figure 1**. (a) Schematic showing the combinatorial sputtering setup employed in this work. The depositions were performed at room temperature (RT) without external heating. (b) Laboratory XRD patterns of the combinatorial Fe-W-N and Fe-Mo-N films as-grown. A vertical offset is applied for clarity.

The effect of post-deposition annealing of the amorphous films was then investigated. Combinatorial strips of Fe-W-N and Fe-Mo-N films were annealed by rapid thermal annealing (RTA) at temperatures ranging from 600 °C to 900 °C for 20 min under flowing $N_2$ (Figure 2a). XRF was performed on post-annealed films; no changes in metal compositions were detected, as expected (Figure S2). High-quality synchrotron GIWAXS patterns of films annealed at 800 °C are shown in Figure 2a and 2b, respectively. The different metal compositions across the graded films result in various phases forming, and the normalized peak intensities for the different phases are plotted versus composition in Figure S4.

First, we discuss the Fe-W-N films (Figure 2b), which exhibit a complex evolution of phases as we tune the composition. At W-rich compositions ($x < 0.5$ where $x$ = Fe/(Fe+$M$) and $M$ = Mo or W), the film mostly exhibits a rocksalt-derived face-centered cubic structure (space group *Fm-3d*), which we refer to as RS in this work. We note the emergence of the body-centered cubic (bcc) phase of W, labelled $\alpha$-W for $x$ = 0.24–0.28. As the material becomes Fe-rich ($x > 0.5$), it rapidly loses the RS structure and forms an $\eta$-nitride phase (nominally $Fe_3W_3N$), as indicated by the characteristic diffraction peaks in GIWAXS patterns. RS + $\eta$ coexistence is observed from $x$ = 0.5–0.61, and then a phase-pure $\eta$ phase in the narrow window $0.61 \leq x \leq 0.64$, corresponding to a composition closer to $Fe_4W_2N$. This deviation from stoichiometry for the phase-pure composition highlights the flexibility of the $\eta$ lattice to accommodate extra Fe atoms. In fact, this effect is well known and has been observed in $\eta$-carbides based on $Fe_3W_3C$ [31,32]. Another study has even reported the bulk synthesis of nitride $Fe_4W_2N$, in which the authors concluded from Rietveld analysis that the excess Fe substitutes onto W sites [33]. While in thin films, detailed site analysis is difficult, our data suggests a compositional tie line over which the $\eta$ phase accommodates Fe excess without losing its structure. We note that since stoichiometric $Fe_3W_3N$ was successfully synthesized in bulk in previous works [20,34], the deviation observed here is unique to our thin-film synthesis approach, in which a metastable RS phase, likely to be $WN_y$, competes with the formation of the $\eta$-nitride at the 1:1 (i.e., $x$ = 0.5) stoichiometry.

In the case of Fe-Mo-N films (Figure 2c), the $\eta$ structure is observed across the whole range of compositions ($x$ = 0.31–0.70). A RS phase emerges at Mo-rich compositions and coexists with the $\eta$ phase. Interestingly, when the film becomes Fe-rich, the bcc phase of Fe ($\alpha$-Fe) is present. Phase-pure $\eta$-nitride is only observed near 1:1 Fe:Mo stoichiometry, in the range $x$ = 0.51–0.59. The single-phase region is correlated with a curious increase in electrical resistivity, from 160 to 260 μΩ·cm, before dropping again at higher Fe content (Figure S3). The persistence of $Fe_3Mo_3N$ $\eta$-based phases at Fe-poor stoichiometry alongside RS or $\alpha$-Fe peaks suggest partial phase separation, unlike in $Fe_3W_3N$, which only forms at Fe-rich compositions. On the other hand, the composition dependence of the lattice parameter in $Fe_3Mo_3N$ shows a slight increase for $x < 0.5$ where RS coexistence is observed, and then a sudden drop around $x$ = 0.5 (Figure S5), which was not observed in bulk studies before. This suggests that a reconfiguration of cation sites in the $\eta$ structure occurs when the Fe content passes the stoichiometric point, provoking an abrupt contraction in the lattice parameter. This notable difference between the two systems indicates that $Fe_3Mo_3N$ exhibits a broader synthesis window than $Fe_3W_3N$ with a greater accommodation for off stoichiometry among secondary phases.

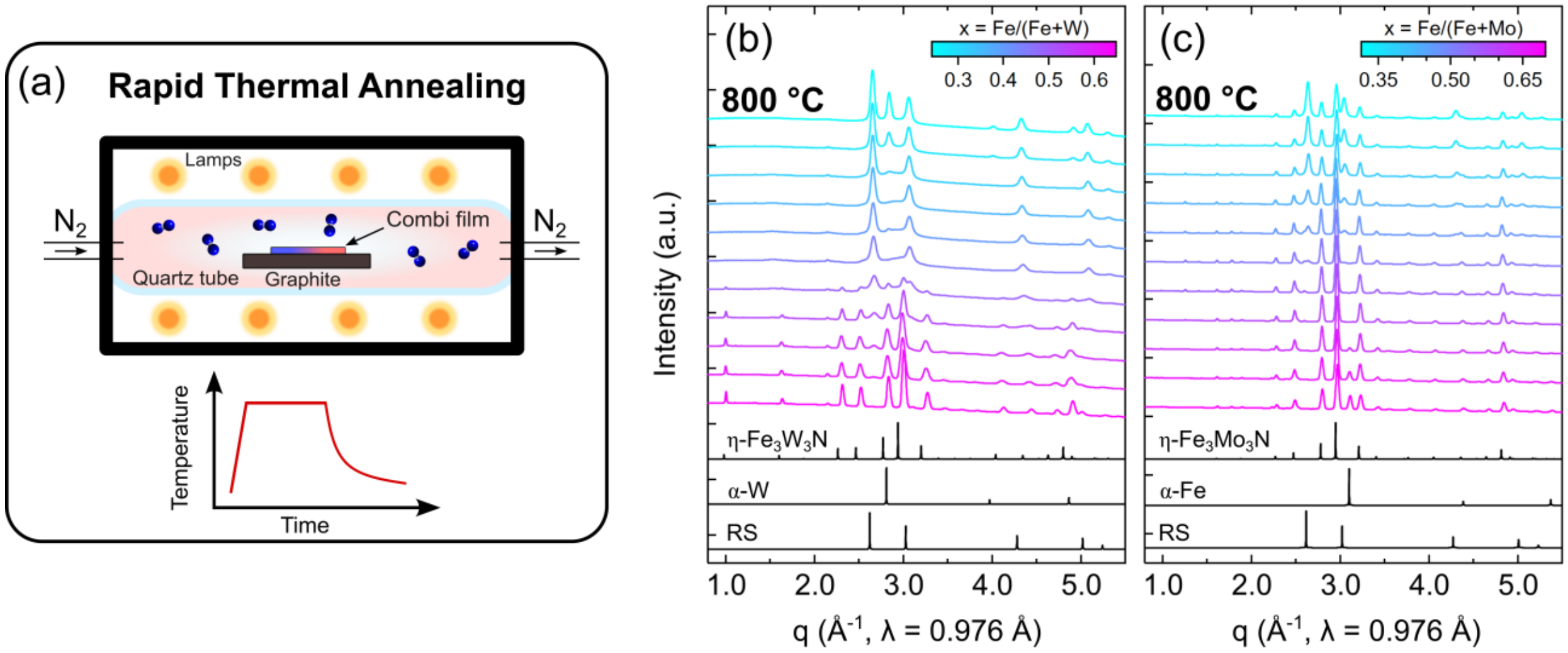

**Figure 2**. (a) Schematic of the rapid thermal annealing setup employed in this work. The bottom graphic illustrates the typical temperature profile employed for the RTA experiments. Synchrotron GIWAXS patterns of (b) Fe-W-N and (c) Fe-Mo-N films annealed by RTA at 800 °C for 20 min ($\lambda$ = 0.976 Å). Simulated patterns of relevant phases are added for reference. The term "RS" refers to rocksalt.

### Thermodynamic Insights into $\eta$-Nitride Stability

***Experimental analysis***. To have a complete picture of the synthesis experiments, experimental phase maps of Fe-W-N and Fe-Mo-N are presented in Figure 3a and 3b, respectively; they summarize the results of XRD / GIWAXS on various combinatorial films annealed from 600 to 900 °C. First, we show that annealing the as-grown amorphous films at 600 °C for 20 min is insufficient to induce crystallization in both systems. For the sample annealed at 700 °C, we observe significant changes for both materials. In Fe-W-N, a rocksalt (RS) phase forms across the whole range of composition x = 0.25 - 0.64, and the $\eta$ phase starts appearing at Fe-rich compositions along with RS, although the crystallinity is mediocre as suggested by low-intensity XRD peaks, not shown here. The existence of RS in the W-rich region likely originates from $WN_y$ which is known to form by sputtering as it is often the case for other sputtered binary nitrides [35–37]. Besides, the composition suggests that a fraction of W is substituted by Fe to form a disordered lattice. Although rocksalt WN is thermodynamically metastable, cation vacancies, nitrogen vacancies, or disorder in general can significantly stabilize it, as is the case for other nitrides and oxides [38–41].

Annealing Fe-Mo-N films at 700 °C yields a more complex landscape of phase mixtures (Figure 3b). Interestingly, $\alpha$-Fe forms at Fe-rich compositions, as opposed to the W-based system. A rocksalt MoN phase is observed around $x = 0.5$ and extends to Mo-rich compositions, suggesting the formation of a similar rocksalt-derived structure as in Fe-W-N. Furthermore, hexagonal $\delta$-MoN, which is the thermodynamic ground state in Mo-N system [42], is identified in the range $x = 0.43$ - $0.68$. Overall, a tie line of phase mixture exists across the composition, and the $\eta$ phase is only observed in a narrow window around $x = 0.39$, along with RS. Those observations suggest that in Fe-rich compositions, Fe and Mo diffuse and segregate to separately form $\delta$-MoN and $\alpha$-Fe. However, the RS phase is more likely to stabilize at higher Mo content, driven by $MoN_y$ formation, and therefore RS predominates in the $x < 0.5$ region. The 800 °C annealing step was discussed above (see Figure 2).

Annealing Fe-W-N films at 900 °C mostly results in the decomposition of the rocksalt phase into bcc W (labelled $\alpha$-W), as it was already observed at high W content at 800 °C. This is likely due to the higher volatilization of N at 900 °C and is in good agreement with previous studies on the annealing of W-N thin films [43,44]. Although the $\eta$ phase exists at Fe-rich compositions, as for 800 °C, it is mixed with $\alpha$-W as well as an additional unknown phase. Therefore, no phase-pure $\eta$ is observed at 900 °C. In the Fe-Mo-N system, annealing at 900 °C results in decomposition of the film into bcc-Mo ($\alpha$-Mo) at Mo-rich compositions. The rapid segregation above 800 °C suggests that excess Fe in the $\eta$ structure is accommodated through kinetic trapping.

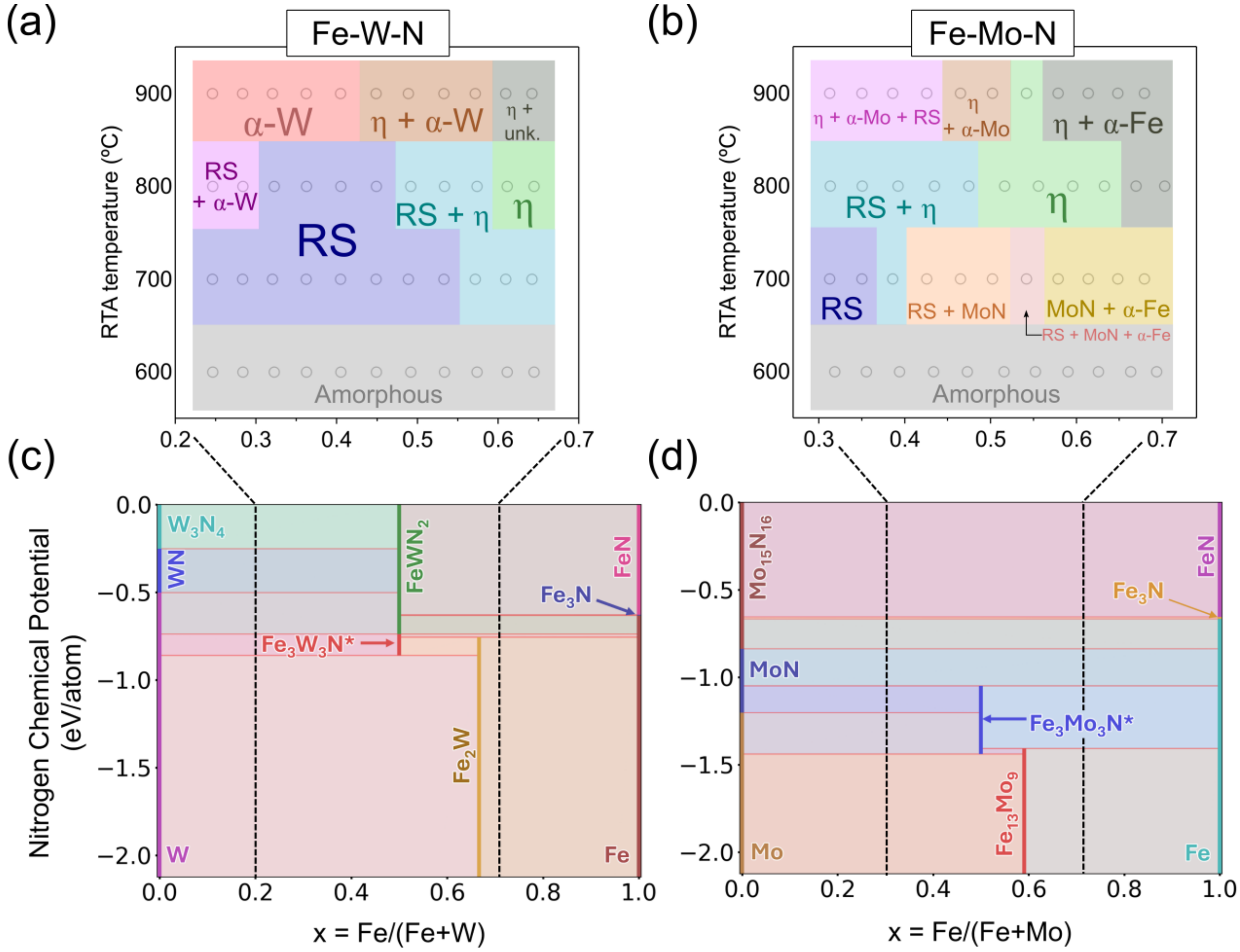

**Figure 3**. (a, b) Experimental phase maps constructed from composition-dependent XRD results Fe-W-N and Fe-Mo-N films respectively, annealed by RTA at temperatures of 600 – 900 °C. The symbol '$\eta$' refers to the $\eta$-nitride phase of nominal formula $Fe_3M_3N$ (with M = W, Mo) and 'RS' stands for rocksalt. (c, d) Nitrogen chemical potential phase diagrams locating the different ground-state phases with respect to metal composition (x-axis) and nitrogen chemical potential (y-axis), with vertical lines representing single phase regions, 2D rectangles representing 2-phase regions and horizontal lines representing 3-phase regions. The dash lines highlight the regions of composition experimentally investigated in this work.

***Computational analysis***. To help understand the differences observed in the two systems, we calculated nitrogen chemical potential ($\mu_N$) phase diagrams for the Fe-W-N and Fe-Mo-N systems; they are presented in Figure 3c and 3d, respectively. This thermodynamic tool illuminates phase stability regions and areas where multiple phases can coexist at equal $\mu_N$. We note that the experimental $\mu_N$ is not trivial to determine as it strongly depends on growth and annealing conditions (including kinetics), so we instead make qualitative comparisons. We also note that the RS phases observed in experiments are not present in the diagrams as stoichiometric WN, MoN or FeN in the rocksalt structure are thermodynamically metastable. The Fe-W-N system contains two stable ternary phases: the $\eta$-nitride $Fe_3W_3N$ and the N-rich hexagonal layered $FeWN_2$, which has previously been prepared [45,46], but were not observed here. The computed phase diagrams also show various $WN_x$ and $FeN_x$ binaries, as well as the intermetallic $Fe_2W$. The Fe-Mo-N diagram is relatively similar with the exception that $FeMoN_2$ (isostructural to $FeWN_2$) does not appear as it is slightly metastable, even though it has been synthesized by ammonolysis [47,48]. Experimentally, we did not synthesize $FeWN_2$, but this is likely because ammonia was not used here during post-deposition annealing.

First, the computational phase analysis shows that the stability window of pure $Fe_3W_3N$ (described by the height in $\mu_N$ of red vertical line at $x = 0.5$) is much narrower in $\mu_N$ than of that of $Fe_3Mo_3N$ (0.1 eV/atom versus 0.4 eV/atom, respectively). This suggests that the former may need more precise synthesis conditions to thermodynamically stabilize. In addition, $Fe_3W_3N$ is stabilized at higher $\mu_N$ (-0.75 eV/atom) than $Fe_3Mo_3N$ (-1.05 eV/atom). In general, higher synthesis temperatures correspond to more negative $\mu_N$, as the chemical potential of gaseous phases scales with $\mu_N = \mu_N^0 + RT\ln(P_{gas}/P^\circ) - TS_{gas}$ where $S_{gas}$ is approximately $\frac{3}{2}nk_BT$ for monatomic gas or $\frac{5}{2}nk_BT$ for diatomic gas. This means that synthesizing $Fe_3Mo_3N$ should be more favorable than $Fe_3W_3N$ at higher temperatures. This is reflected in the experimental phase maps, as the $Fe_3Mo_3N$ can be retained in phase-pure form at 900°C, whereas the $Fe_3W_3N$ was not observed in phase-pure form at higher temperatures. The difference in these stabilizing nitrogen chemical potentials arises from the difference in formation energy between WN (-0.251 eV/atom) and MoN (-0.602 eV/atom) [49].

$Fe_3Mo_3N$ is computed to have a 2-phase coexistence with both elemental Fe and Mo over an appreciable $\mu_N$ window as described by the vertical width of [Fe + $Fe_3Mo_3N$] and [Mo + $Fe_3Mo_3N$] rectangles of 0.36 and 0.24 eV/atom, respectively. In contrast, $Fe_3W_3N$ has a small 2-phase coexistence window with W, as described by the width of the [W + $Fe_3W_3N$] rectangle in red, as well as an extremely narrow 2-phase coexistence with Fe, as described by the very thin rectangle [Fe + $Fe_3Mo_3N$] above $Fe_2W$. This is in strong agreement with the experimental maps in which a large coexistence of $Fe_3Mo_3N$ with both $\alpha$-Fe or $\alpha$-Mo was observed, as opposed to $Fe_3W_3N$ in which coexistence with $\alpha$-W exists around $x = 0.5$ but no coexistence with $\alpha$-Fe was observed, in the detection limits of laboratory XRD. These findings suggest that the synthesis of $Fe_3W_3N$ requires greater compositional and chemical potential control.

The decomposition into $\alpha$-W or $\alpha$-Mo at high temperature can be understood as at fixed pressure, the experimental $\mu_N$ is expected to decrease with temperature. However, since both $Fe_3W_3N$ and $Fe_3Mo_3N$ persist at 900 °C, this could indicate that $\mu_N$ is still large enough to stabilize the $\eta$ phases. We note that in Fe-W-N, no phases located above $Fe_3W_3N$ in $\mu_N$ are observed in the experiments, as opposed to Fe-Mo-N in which hexagonal MoN is obtained at 700 °C and then vanishes at 800 °C to favor $Fe_3Mo_3N$, and implies that $\mu_N$ falls below -1.30 eV where MoN window ends.

### Crystallographic Analysis of $\eta$-Nitride Films

To further characterize the crystal structure of the $\eta$-nitride phase in films annealed at 800 °C, we performed refinements of GIWAXS patterns using the LeBail method at compositions where the $\eta$ phase is present without distinct secondary phases and shows maximum peak intensities in XRD (Figure S4). Thus, we selected $Fe_{3.84}W_{2.16}N$ (corresponding to $x = 0.64$) and $Fe_{3.54}Mo_{2.46}N$ (i.e., $x = 0.59$). Results of the fits are presented in Figure 4a and 4b. The corresponding detector images converted to reciprocal space are displayed in Figure 4c and 4d.. The good continuity of diffraction rings confirms that both films are polycrystalline without any preferential orientations. The GIWAXS patterns were fitted to the $\eta$-nitride structure of $Fe_3W_3N$ and $Fe_3Mo_3N$ in space group *Fd-3m* (Figure 4e).

For $Fe_{3.84}W_{2.16}N$ (Figure 4a), the LeBail refinement yields a lattice parameter $a$ = 10.8556(2) Å with a profile weighted R-factor $R_{wp}$ = 1.95%. It is interesting to note that the lattice parameter is significantly smaller than the value reported in literature for $Fe_3W_3N$ powder prepared by ammonolysis ($a$ = 11.11 Å) [34]. This could be due to relaxation occurring during rapid annealing or simply due to the Fe-rich stoichiometry, which is closer to $Fe_4W_2N$ in this case. In fact, the lattice parameter of $Fe_4W_2N$ powder has already been reported and is slightly smaller than $Fe_3W_3N$ ($a$ = 11.08 Å) [50]. However, this reduction by only 0.3% from $Fe_3W_3N$ to $Fe_4W_2N$ in bulk materials cannot explain the significant difference we observe in our thin-film. Another possibility could be due to the nitrogen stoichiometry. Many materials are also stable in the $\eta$ structure with N-deficient 6:6:1 stoichiometry and a smaller lattice parameter, such as $Fe_6W_6C$ or $Fe_6W_6N$ (Figure S8). [32,51,52]. The absence of peak splitting in our GIWAXS patterns rules out the coexistence of both $Fe_3W_3N$ and $Fe_6W_6N$ phases. Despite the good phase purity, we note the presence of a low-intensity peak at $q$ = 3.1 $Å^{-1}$ (magnified in Figure S6), corresponding to (110) reflection of $\alpha$-Fe, which was not detected by laboratory XRD. Two additional low-intensity peaks are observed at low $q$, marked with asterisks (*), suggesting that other small crystalline impurities are present. The low amplitude of these peaks, relative to the main peaks, could suggest that these are limited to the surface and could come from oxides formed during annealing, especially due to the excess of Fe. However, we could not index those peaks to any oxide phase with certainty.

The refinement of $Fe_{3.54}Mo_{2.46}N$ (*x* = 0.59) shown in Figure 4b yields a lattice parameter $a$ = 11.0277(2) Å with $R_{wp}$ = 2.36%. The value of $a$ is close to the reported value for bulk $Fe_3Mo_3N$, although slightly smaller ($a$ = 11.067 Å) [53]. This difference is likely due to the excess Fe, which reduces the lattice size, as discussed previously. Although the fit is of good quality and shows decent phase purity, a few low-intensity peaks from unknown phases are detected at low $q$ (*) and are likely to come from surface oxides, as in the W system, but do not match with $\alpha$-Fe. In conclusion, these refinements confirm that both $Fe_{3.84}W_{2.16}N$ and $Fe_{3.54}Mo_{2.46}N$ film show good phase purity of $\eta$-nitride, with respect to a small amount of unknown impurities which we associate with surface oxides. Nevertheless, these films can still be considered "phase-pure" in bulk.

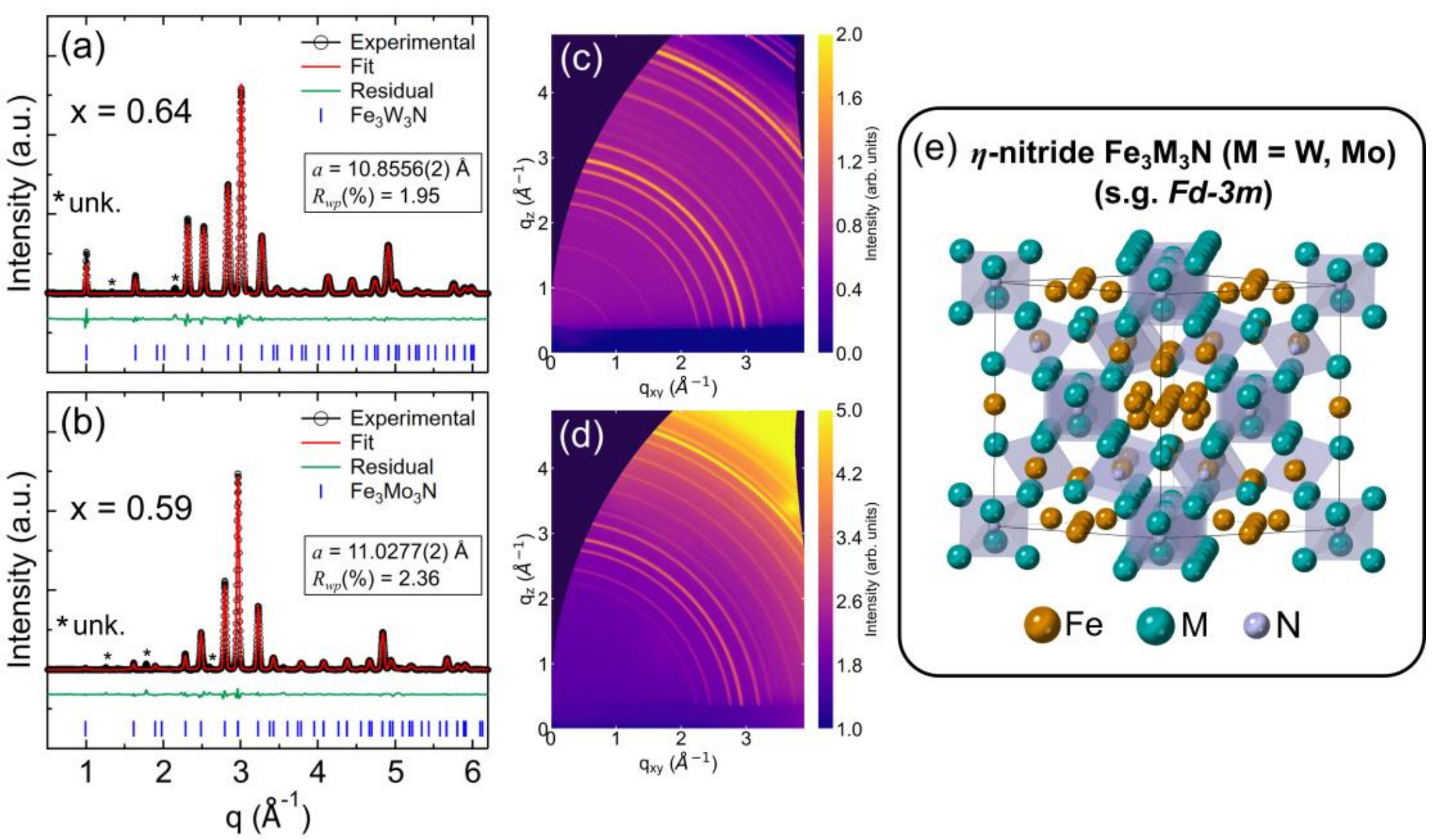

**Figure 4**. (a, b) LeBail refinements of synchrotron GIWAXS data and (c, d) the corresponding GIWAXS detector images converted into reciprocal-space of (a, c) $Fe_{3.84}W_{2.16}N$ ($x$ = 0.64) and (b, d) $Fe_{3.54}Mo_{2.46}N$ ($x$ = 0.59). The label "unk." refers to unknown phases. (e) Crystal structure of $\eta$-nitride $Fe_3M_3N$ ($M$ = W, Mo).

## Dependance of Ferromagnetism upon Subtle Composition Changes

The $\eta$-carbide type compounds containing magnetic elements are known to exhibit interesting and complex itinerant-type magnetism in bulk form arising from the geometrical frustration inherent to the $\eta$-carbide type structure, and the delicate balance between metal-metal and metal-nitrogen bonding which makes these systems highly responsive to changes in composition [54,55]. Thin films offer an additional degree of control over composition, disorder, and defect concentration, which may further tune magnetic interactions. Motivated by these considerations, we investigate in the following sections how deviations from stoichiometry and phase competition influence the magnetic response of $Fe_3W_3N$ and $Fe_3Mo_3N$ films annealed at 800 °C. We compare Fe-rich compositions exhibiting phase-pure $\eta$ as observed by GIWAXS, with near-stoichiometric films ($x$ ~ 0.5) where secondary RS phases are present (as highlighted in Figure S7).The magnetic data presented in the following sections are normalized per formula unit of $Fe_3W_3N$ or $Fe_3Mo_3N$ to aid in analysis at low temperature.

***Magnetic susceptibility***. Figure 5 shows the temperature-dependent magnetic DC susceptibility $\chi = M/H$ (where $M$ is the magnetization and $H$ the applied field) measured under zero-field cooled (ZFC) and field-cooled (FC) conditions with a constant field of 5 kOe for $\eta$-nitride films at Fe-rich composition as well as near stoichiometry. Both systems display Curie-Weiss-like paramagnetic behavior at high temperature with a monotonic decrease of $\chi$ upon warming. In the Fe-W-N system (Figure 5a), the susceptibility shows a smooth transition around 130 K at both compositions, indicating the onset of ferromagnetic (FM) ordering, which aligns with the observations reported in bulk $Fe_3W_3N$ [20]. The Fe-rich sample $Fe_{3.84}W_{2.16}N$ ($x$ = 0.64) shows a significant vertical offset at high temperature, which can be ascribed to FM impurities with a high Curie temperature, most likely coming from the $\alpha$-Fe secondary phase (see Figure S6). In the near-stoichiometric sample $Fe_{3.12}W_{2.82}N$ ($x$ = 0.53), the slightly negative $\chi$ at high temperature could be due to a weak diamagnetic background, even after substrate subtraction.

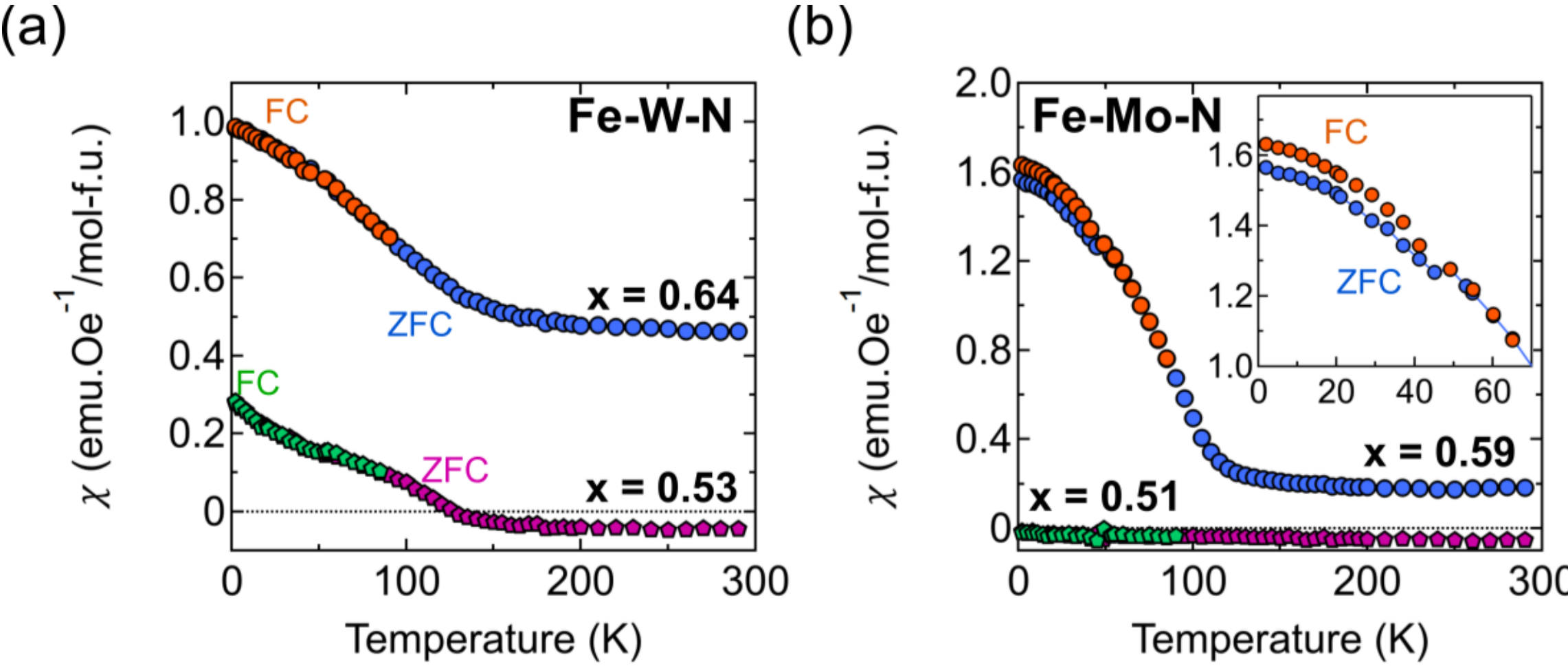

**Figure 5**. Temperature-dependent magnetic susceptibility of $\eta$-nitride films (a) $Fe_{3.84}W_{2.16}N$ (x = 0.64) and $Fe_{3.18}W_{2.82}N$ (x = 0.53), and (b) $Fe_{3.54}Mo_{2.46}N$ (x = 0.59) and $Fe_{3.06}W_{2.94}N$ (x = 0.51) . The susceptibility was measured while heating up from 2 K, under an applied field of 5 kOe (0.5 T) in ZFC and FC conditions. The FC traces are only displayed at low temperature for clarity. In the case of $Fe_{3.54}Mo_{2.46}N$, the inset magnifies the small ZFC/FC splitting at low temperature. Although the compositions differ in stoichiometry, the magnetic susceptibility is normalized per mole of formula unit of $Fe_3W_3N$ and $Fe_3Mo_3N$, respectively.

The Fe-Mo-N system displays qualitatively different behavior (Figure 5b). In $Fe_{3.54}Mo_{2.46}N$ ($x$ = 0.59), a pronounced transition in $\chi$ is observed around the same transition temperature as in Fe-W-N. The susceptibility sharply increases and seems to saturate around 1.6 emu.Oe$^{-1}$/mol-f.u. at 2 K, which is much higher than in the W system and indicates a more pronounced FM ordering. This FM transition is accompanied by a small bifurcation between ZFC and FC susceptibility curves below approximately 50 K (Figure 5b inset). While subtle, this splitting could suggest the presence of competing interactions at low temperature, potentially due to site disorder, domain pinning, or coupling between multiple magnetic phases. Although no crystalline $\alpha$-Fe is detected by GIWAXS at this composition, the small offset in $\chi$ (~0.2 emu.Oe$^{-1}$/mol-f.u.) at high temperature suggests a minor ferromagnetic impurity, potentially amorphous or nanocrystalline Fe. As opposed to Fe-W-N, the near-stoichiometric film $Fe_{3.06}Mo_{2.94}N$ ($x$ = 0.51) shows no magnetic ordering across the entire temperature range studied, consistent with prior neutron diffraction studies on bulk samples [18], and confirming that FM behavior in this system is induced by tuning the stoichiometry, rather than intrinsic to the $Fe_3Mo_3N$ $\eta$-nitride structure.

***Field-dependent magnetization***. Figure 6 shows the field-dependent magnetization loops at different temperatures. The $Fe_{3.06}Mo_{2.94}N$ film is not shown as it shows no magnetic ordering. In Fe-W-N, the presence of hysteresis in both samples confirms the FM ordering. In $Fe_{3.18}W_{2.82}N$ (Figure 6a), the hysteresis vanishes beyond 100 K, and the magnetization becomes slightly negative, likely an artefact stemming from low signal and substrate subtraction. This corroborates the absence of an $\alpha$-Fe impurity at this near-stoichiometric composition, illustrated by the saturation magnetization ($M_s$) and the coercive field ($H_c$) reaching zero at high temperature (Figure 7a and 7c). $H_c$ and $M_s$ linearly increases below a Curie temperature ($T_c$) estimated at 130 K and respectively reach 223 Oe and 0.27 $\mu_B$/f.u. at 2 K. In $Fe_{3.84}W_{2.16}N$, the hysteresis persists at 300 K, confirming that $\alpha$-Fe ferromagnetic impurity dominates at high temperature. We note that the normalization per formula unit of $Fe_3W_3N$ is not valid in this temperature range. The emergence, temperature dependence, and disappearance of hysteresis near the ordering transition indicate that the dominant low-temperature magnetic behavior is associated with the $\eta$-nitride phase. The shape of the hysteresis loop (or “squareness”) is more pronounced below 150 K, suggesting that the FM ordering of the $\eta$ phase dominates over the impurity. The coercive field (Fig 7a) is 137 Oe at 2 K and decreases with temperature as expected, before a sudden jump to 232 Oe around the transition where FM from $\alpha$-Fe takes over, which is in the typical range of coercivity observed in $\alpha$-Fe thin films after annealing [56]. The magnitude of $H_c$ in the $\eta$ phase is relatively small and suggests the material is a soft ferromagnet.

In Figure 6c, $Fe_{3.54}Mo_{2.46}N$ shows clear hysteresis with greater squareness than Fe-W-N films. The hysteresis region shrinks with temperature until completely vanishing beyond 100 K, confirming the FM transition observed in susceptibility measurements. The coercive field at 2 K is measured at 107 Oe and decreases with temperature within the FM regime, as shown in Figure 7b. Moreover, it shows a two-step trend by first reaching a plateau around 50 K and then rapidly drops to zero while crossing the FM transition upon cooling. This unusual behavior could indicate the presence of two magnetic populations. The $M_s$ extracted at higher field reaches 1.37 $\mu_B$/f.u. (0.46 $\mu_B$/Fe), assuming the formula unit of $Fe_3Mo_3N$, and its temperature dependance, shown in Figure 7, aligns well with the transition observed in the magnetic susceptibility. Since the stoichiometric film does not show FM, it is interesting to estimate the magnetization per excess of Fe in $Fe_{3.54}Mo_{2.46}N$. If we subtract the offset at 300 K (0.14 $\mu_B$/f.u.) due to impurity, $M_s$ at 2K becomes 1.23 $\mu_B$/f.u. Since there are 0.54 Fe atoms in excess per formula unit, we obtain $M_s$ = 2.28 $\mu_B$ per excess Fe atoms, which is below the solubility limit determined in the $Fe_3W_3N$ system. This value is close to the magnetic moment of $\alpha$-Fe (2.2 $\mu_B$), suggesting some itinerant magnetism.

Overall, the $\eta$-nitride films investigated here exhibit soft ferromagnetic behavior with a Curie temperature around 130 K, characterized by modest magnetization and low coercivity (on the order of ~100 Oe at low temperature). These values are significantly smaller than those of high-performance nitride-based candidates for permanent magnets such as $Fe_{16}N_2$ or Sm-Fe-N compounds, which typically display substantially larger magnetization and coercivity [57,58]. More interestingly, while $Fe_3Mo_3N$ shows no magnetic ordering, modest excess of Fe induces itinerant-type ferromagnetism while maintaining the $\eta$ structure. Therefore, the magnetic softness and great sensitivity to composition highlight a different opportunity: $\eta$-nitrides serve as a platform for investigating composition-driven magnetic instabilities and frustration in nitrogen-poor transition-metal systems. In this context, their value lies less in maximizing magnetic performance and more in enabling controlled exploration of correlated and emergent magnetic behavior in structurally robust nitride frameworks.

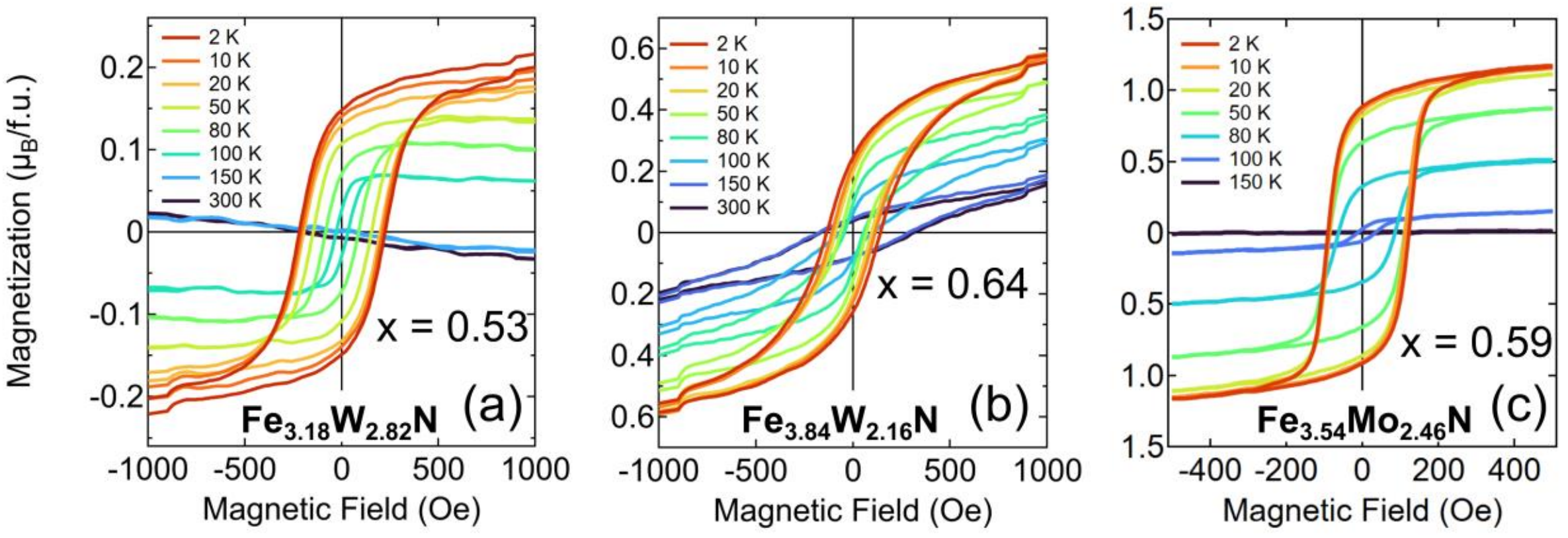

**Figure 6**. Magnetization loops of selected $\eta$-nitride films at different temperatures: (a) $Fe_{3.18}W_{2.82}N$ (x = 0.53), (b) $Fe_{3.84}W_{2.16}N$ (x = 0.64), and (c) $Fe_{3.54}Mo_{2.46}N$ (x = 0.59).

### Exchange-Bias Type Behavior in $Fe_{3.54}Mo_{2.46}N$

A small shift in the hysteresis loop is observed in $Fe_{3.54}Mo_{2.46}N$ at low temperature, as highlighted in Figure 7e, suggesting exchange-bias (EB) type behavior with an EB field ($H_{EB}$) of approximately 16 Oe at 2 K. While small in magnitude, this effect systematically changes with temperature and is reproducible. Moreover, such effect was not observed in comparable $Fe_{3.84}W_{2.12}N$ films. EB typically arises from interfacial coupling between different magnetic regions, such as FM, AFM, and magnetically disordered phases [59–61]. In the present case, the origin of this EB behavior cannot be unambiguously identified. It may reflect magnetic heterogeneity associated with compositional disorder, nanoscale phase separation, or competing magnetic interactions within the Fe-rich $\eta$-nitride stability window.

Although bulk $Fe_3Mo_3N$ was first categorized as antiferromagnetic, following studies confirmed the absence of magnetic long-range order down to 10 K [18,53]. The same authors showed that $Fe_3Mo_3N$ is located near a quantum critical point and that FM order can easily be induced by small chemical substitutions of ferrous elements such as Co [18,22]. Our observation that $Fe_{3.06}Mo_{2.94}N$ films remain non-magnetic while modest excess Fe induces FM is consistent with this picture. In this context, the EB-like response observed only in Fe-rich films may reflect magnetic heterogeneity arising from proximity to this instability, which is enabled by off-stoichiometric compositions accessible to our thin-film synthesis approach. In thin films, modest Fe excess may be accommodated either through substitutional disorder or nanoscale phase heterogeneity within the $\eta$-nitride stability window. While the microscopic origin of the induced ferromagnetism and exchange-bias-like response cannot be resolved from the present data, these results highlight the pronounced sensitivity of $Fe_3Mo_3N$ magnetism to small deviations from stoichiometry.

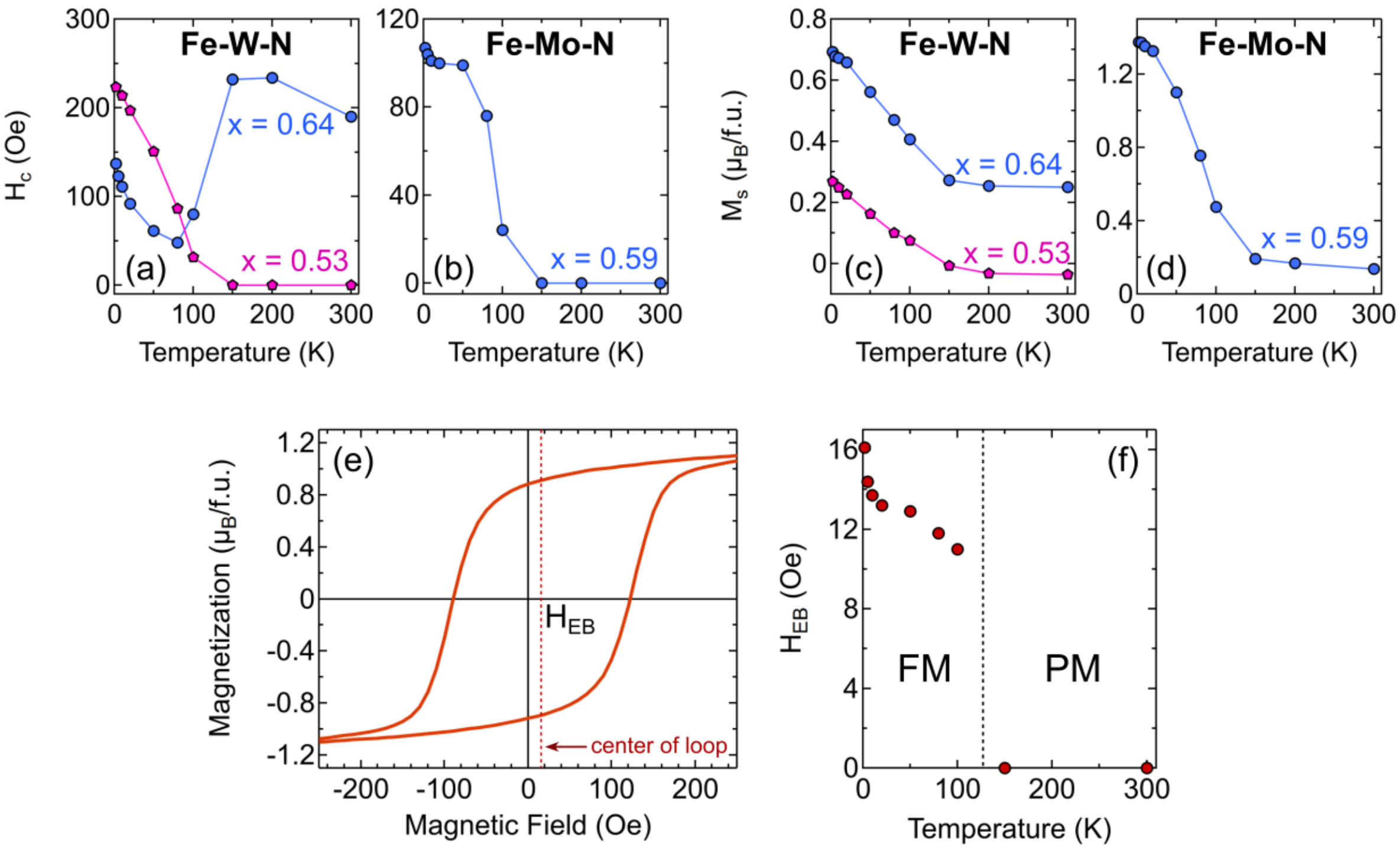

**Figure 7**. Temperature-dependent coercive field ($H_c$) and saturation magnetization ($M_s$) of (a, c) $Fe_{3.84}W_{2.16}N$ (x = 0.64) and $Fe_{3.18}W_{2.82}N$ (x = 0.53) and (b, d) $Fe_{3.54}Mo_{2.46}N$ (x = 0.59). $M_s$ values are extracted at 10 kOe (see Figure S10). (e) Magnetization loop of $Fe_{3.54}Mo_{2.46}N$ recorded at 2 K and highlighting the horizontal shift of the loop. $H_{EB}$ refers to the exchange-bias (EB) field. (f) Temperature dependence of $H_{EB}$ in $Fe_{3.54}Mo_{2.46}N$. FM and PM refer to ferromagnetic and paramagnetic regions, respectively.

## Conclusion

In summary, using a combinatorial synthesis approach and post-deposition RTA in the 600 – 900 °C range, we demonstrate that $\eta$-nitride $Fe_3W_3N$ and $Fe_3Mo_3N$ can be stabilized as thin films, and we identify synthetic windows to achieve phase-pure products. While $\eta$-$Fe_3Mo_3N$ forms across a wide composition range, $\eta$-$Fe_3W_3N$ is only obtained at Fe-rich compositions. Thermodynamic calculations showing significant differences in stability windows between the Fe-Mo-N and Fe-W-N systems rationalize the more precise synthesis conditions needed to synthesize thin-films of $Fe_3W_3N$. A sharp lattice parameter contraction beyond stoichiometry in $Fe_3Mo_3N$ indicates a structural response to excess Fe, pointing to either a rearranged Fe-rich $\eta$ phase or nanoscale phase separation within the $\eta$-nitride stability field. Magnetic measurements on films reveal ferromagnetic ordering in $Fe_3W_3N$ around $T_C$ ~ 130 K. Interestingly, modest excess of Fe induces clear FM in $Fe_3Mo_3N$ around the same temperature, as well as a small exchange-bias effect, suggesting competition between different magnetic phases or sublattices. These findings emphasize the sensitivity of $\eta$-nitrides to off-stoichiometry, reveal their phase stability and structural flexibility in thin-film form, and demonstrate how careful synthetic control enables access to emergent magnetic behavior in nitrogen-poor nitrides.

## Acknowledgements

This work was authored by the National Laboratory of the Rockies (NLR), operated by Alliance for Energy Innovation, LLC, for the U.S. Department of Energy (DOE) under Contract No. DE-AC36-08GO28308. Funding was provided by the U.S. Department of Energy, Office of Science, Basic Energy Sciences, Division of Materials Science, through the Office of Science Funding

Opportunity Announcement (FOA) Number DE- FOA-0002676: Chemical and Materials Sciences to Advance Clean-Energy Technologies and Transform Manufacturing. Use of the Stanford Synchrotron Radiation Lightsource, SLAC National Accelerator Laboratory, is supported by the U.S. Department of Energy, Office of Science, Office of Basic Energy Sciences under Contract No. DE-AC02-76SF00515. This research also used the Theory and Computation facility of the of the Center for Functional Nanomaterials (CFN), which is a U.S. Department of Energy Office of Science User Facility, at Brookhaven National Laboratory under Contract No. DE-SC0012704. The views expressed in the article do not necessarily represent the views of the DOE or the U.S. Government.

## For Table of Contents Only

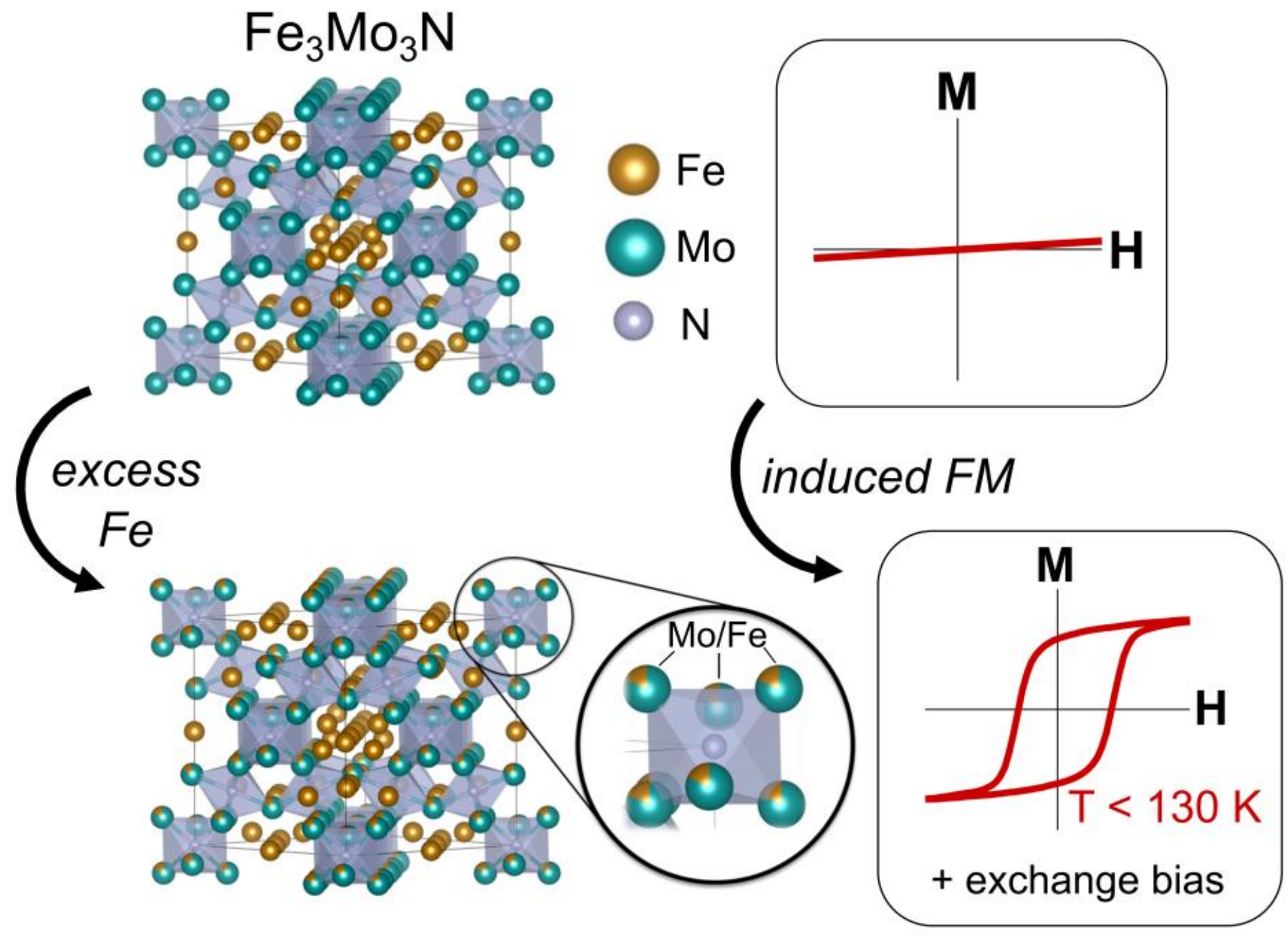

# Thin-Film Stabilization and Magnetism of $\eta$-Carbide Type Iron Nitrides

## Supplementary Materials

Baptiste Julien[1]*, Abrar Rauf[2], Liam Nagle-Cocco[3], Rebecca W. Smaha[1], Wenhao Sun[2], Andriy, Zakutayev[1], Sage R. Bauers[1]*

## I. X-ray Fluorescence (XRF)

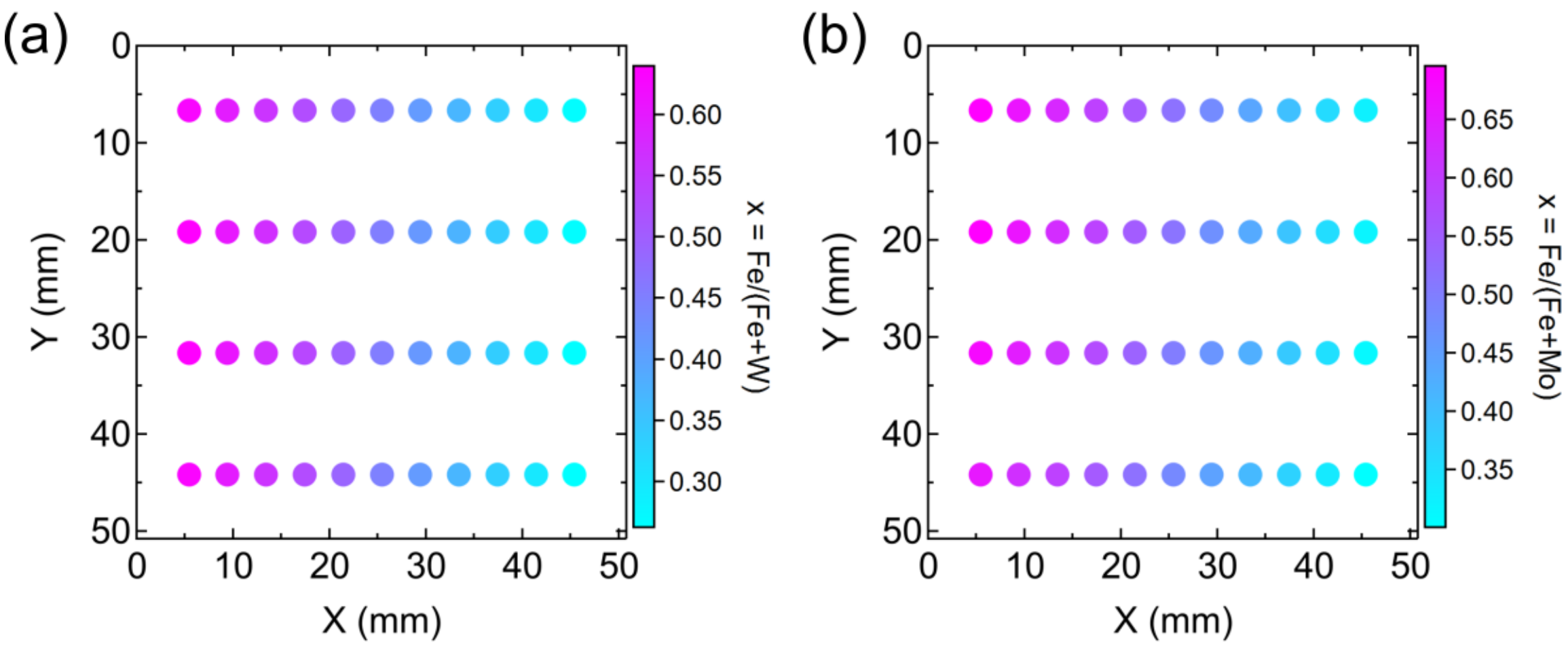

**Figure S1**. XRF maps of as-grown combinatorial 2” libraries of (a) Fe-W-N and (b) Fe-Mo-N. The metal atomic ratio x = Fe/(Fe+*M*) (*M* = W, Mo) is extracted at each position on the 2” library which is divided into 4 rows and 11 columns (44 points total). In the confocal setup used in this work, a horizontal composition gradient is obtained whereas the vertical gradient is minimal. Thus, each row can be cleaved and used separately for specific experiments.

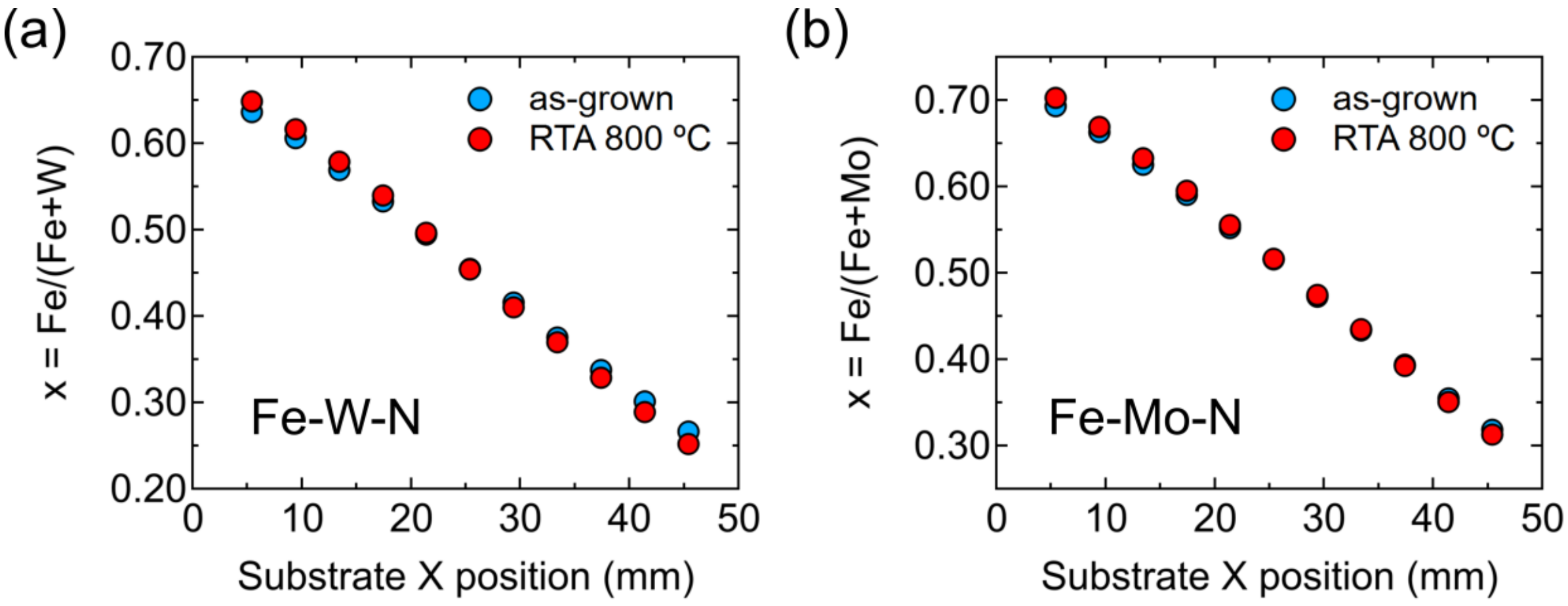

**Figure S2**. Comparison of metal composition between as-grown and annealed films in (a) Fe-W-N and (b) Fe-Mo-N. The difference is minimal, showing that no horizontal diffusion occurs during annealing.

## II. Electrical Resistivity

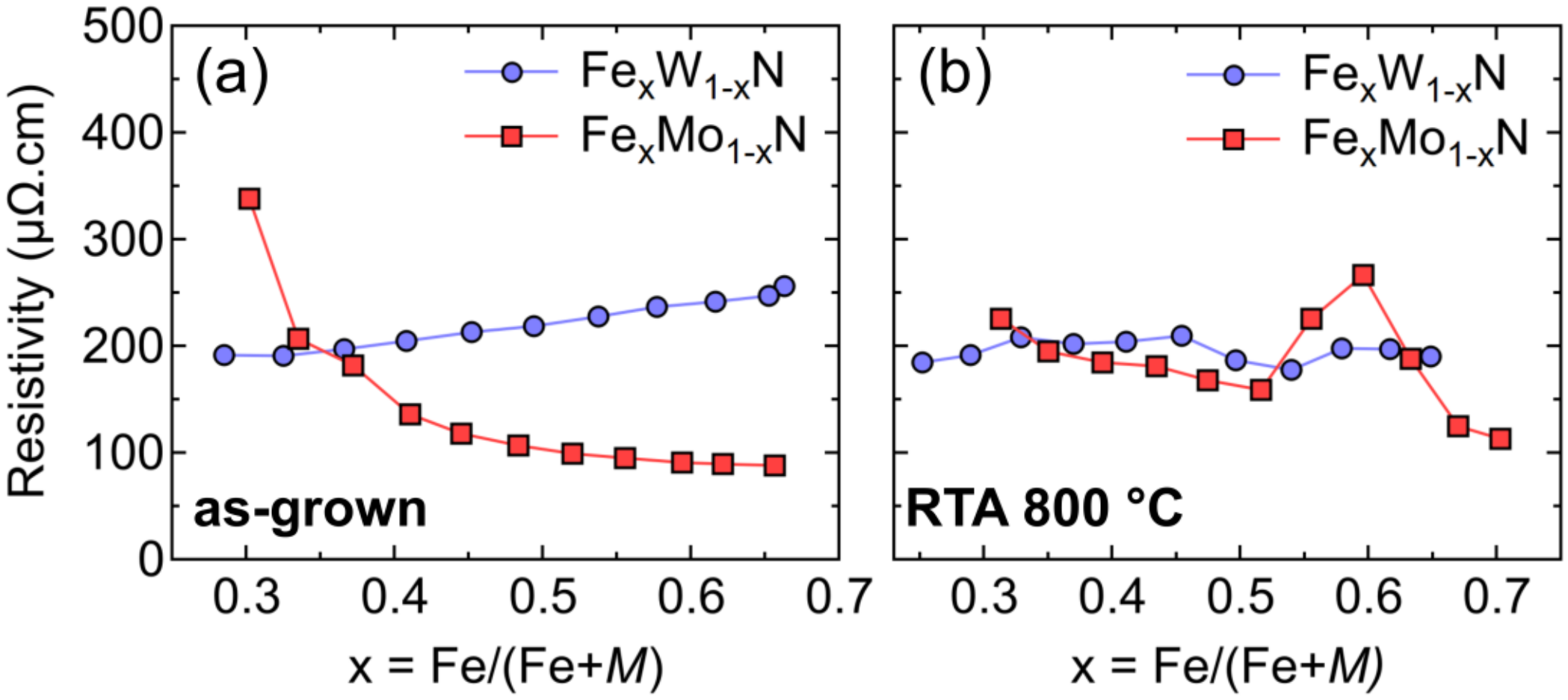

**Figure S3**. Composition-dependent resistivity of (a) as-grown $Fe_xW_{1-x}N$ and $Fe_xMo_{1-x}N$ amorphous films and (b) films annealed at 800 °C. The resistivity was calculated from the sheet resistance measured by four-point probe method and from the local thickness extracted from XRF.

## III. Structural Characterization

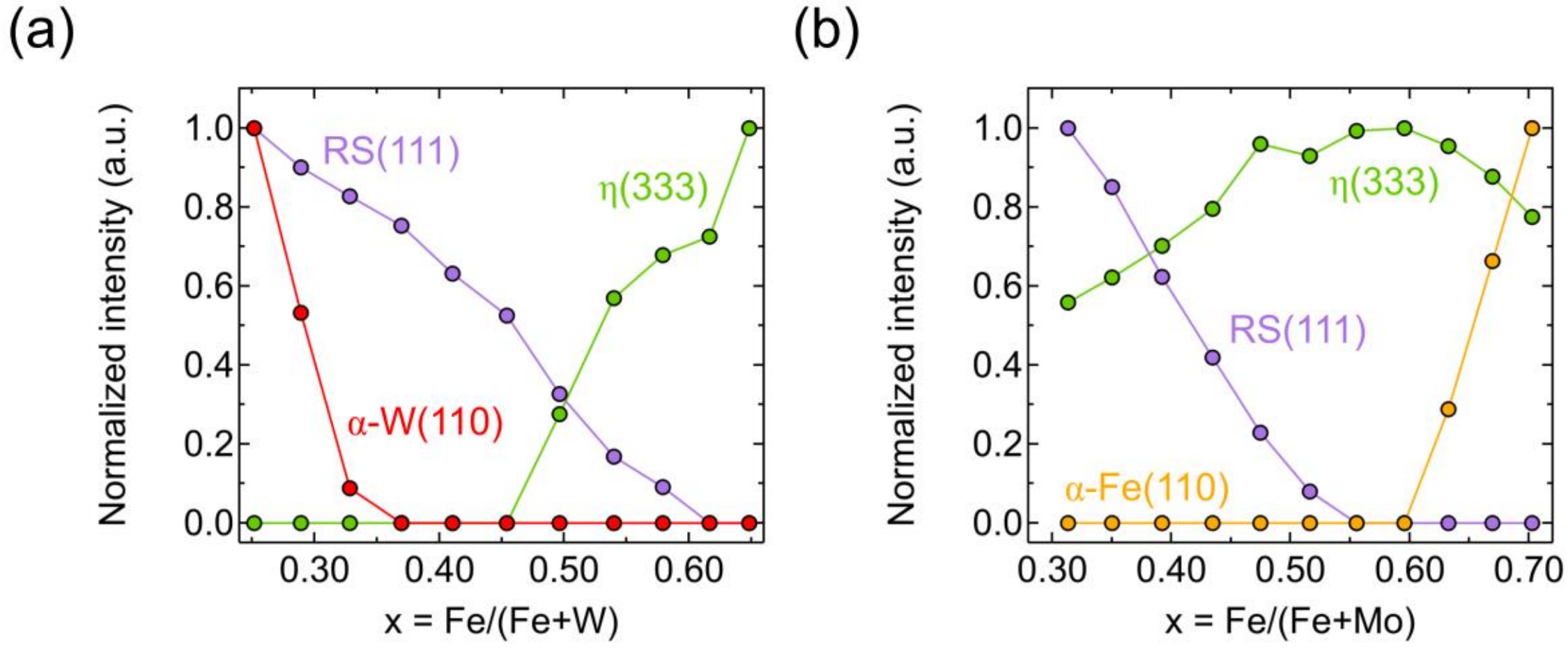

**Figure S4**. Evolution with composition of XRD peak normalized intensities of the different phases identified in (a) $Fe_xW_{1-x}N$ and (b) $Fe_xMo_{1-x}N$ films annealed at 800 °C. The labels '$\eta$' and 'RS' refer to the $\eta$-nitride phase and the rocksalt phase, respectively.

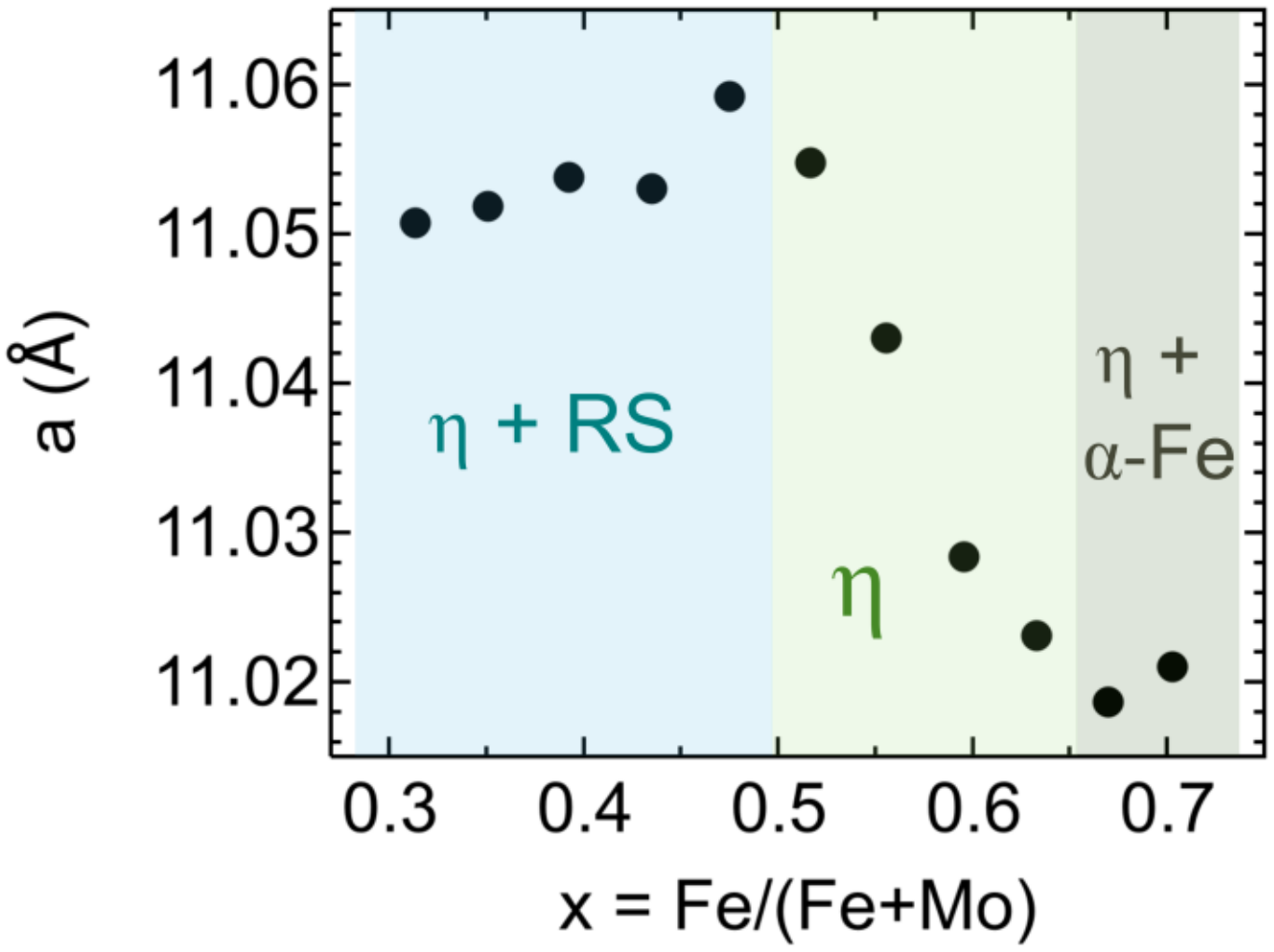

**Figure S5**. Composition dependance of the $\eta$ lattice parameter in $Fe_xMo_{1-x}N$ film annealed at 800 °C. A sudden drop in the lattice parameter occurs in the phase-pure region.

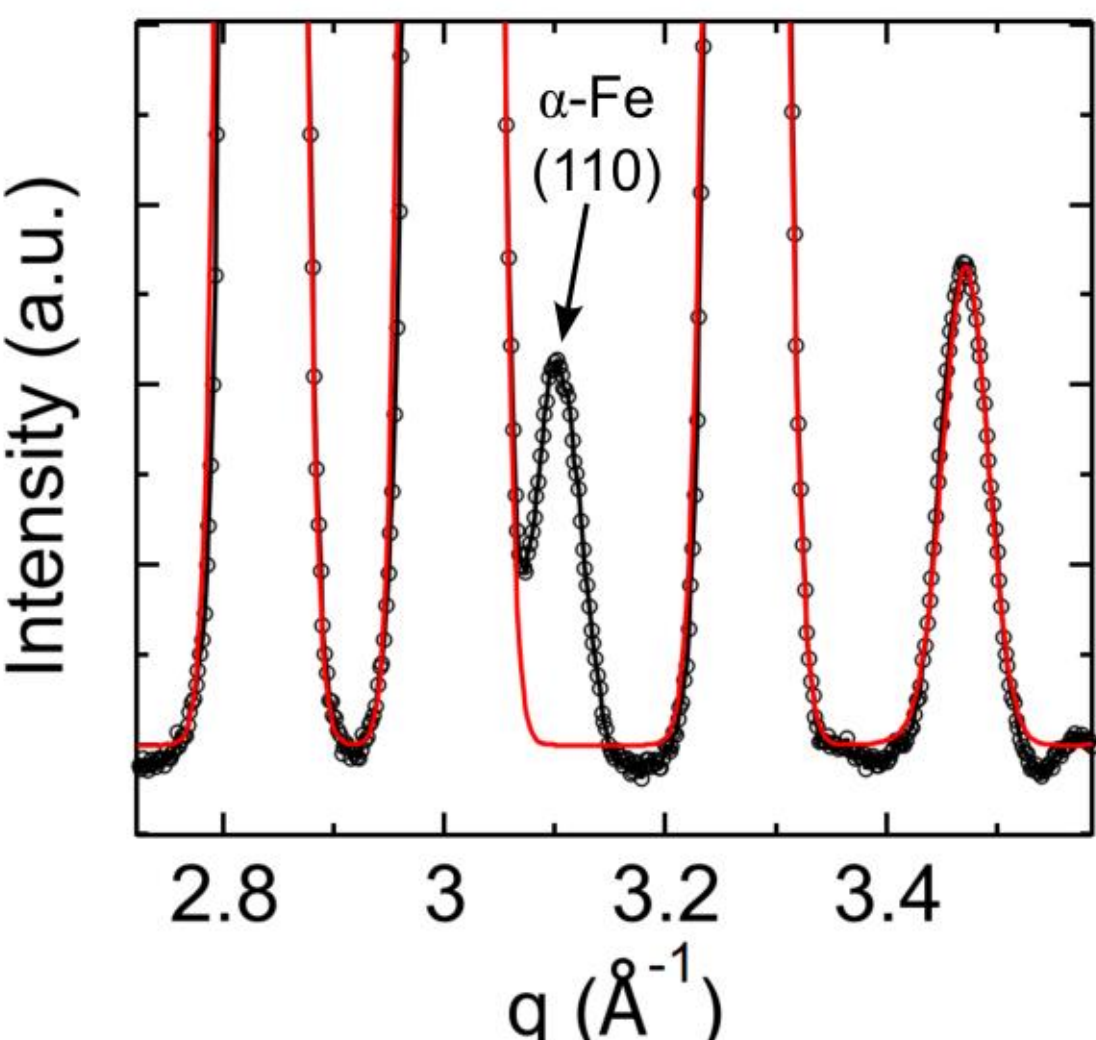

**Figure S6**. GIWAXS patterns of $\eta$-$Fe_3W_3N$ and LeBail fit magnifying around the low-intensity reflection at $q$ = 3.1 Å$^{-1}$ corresponding (110) of $\alpha$-Fe secondary phase.

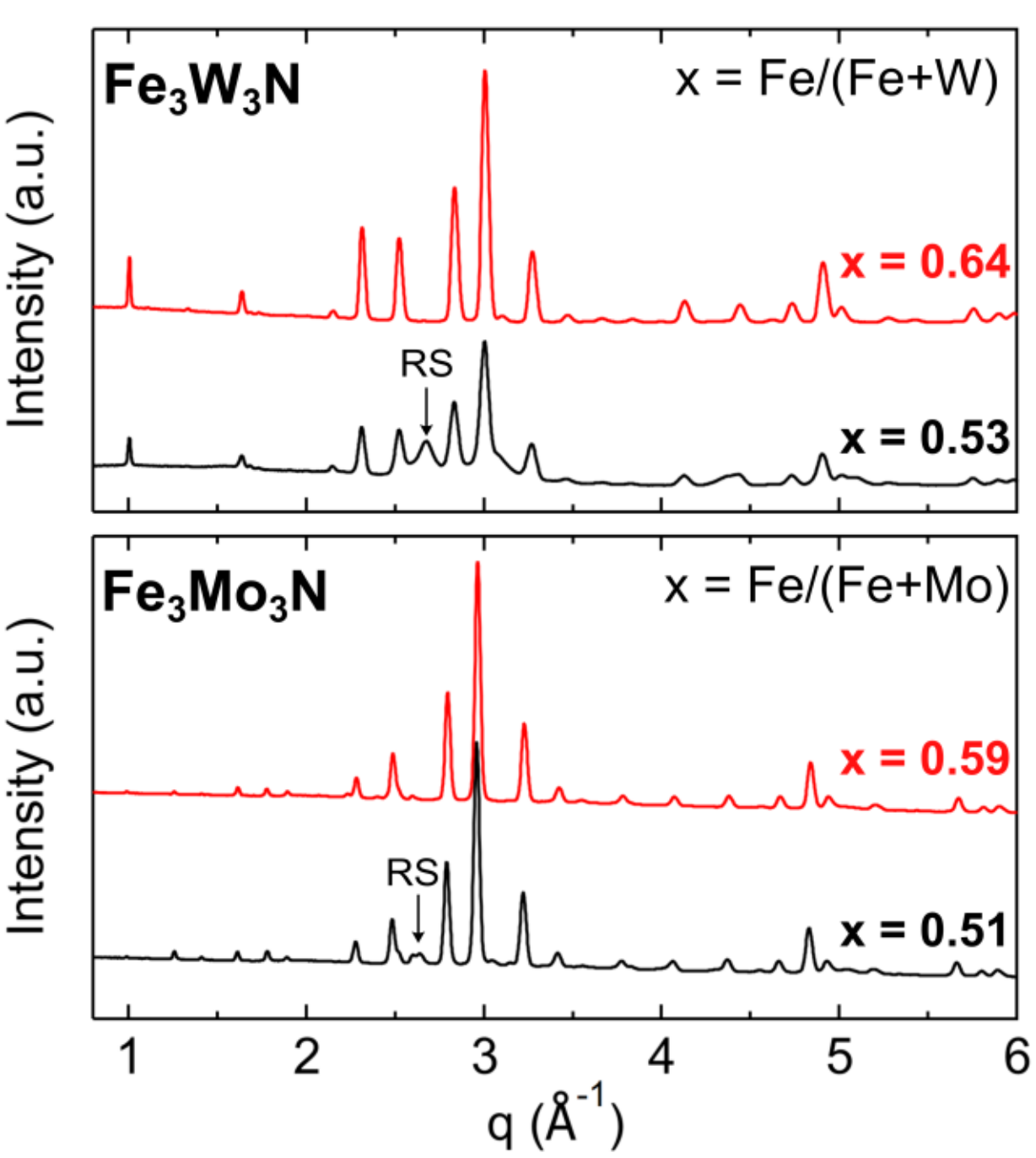

**Figure S7**. GIWAXS patterns of $Fe_xW_{1-x}N$ and $Fe_xMo_{1-x}N$ films annealed at 800 °C comparing two different metal-to-metal ratios: single-phase $\eta$ at Fe-rich compositions (x > 0.5) and near-stoichiometry (x ~ 0.5) which exhibit rocksalt (RS) impurities.

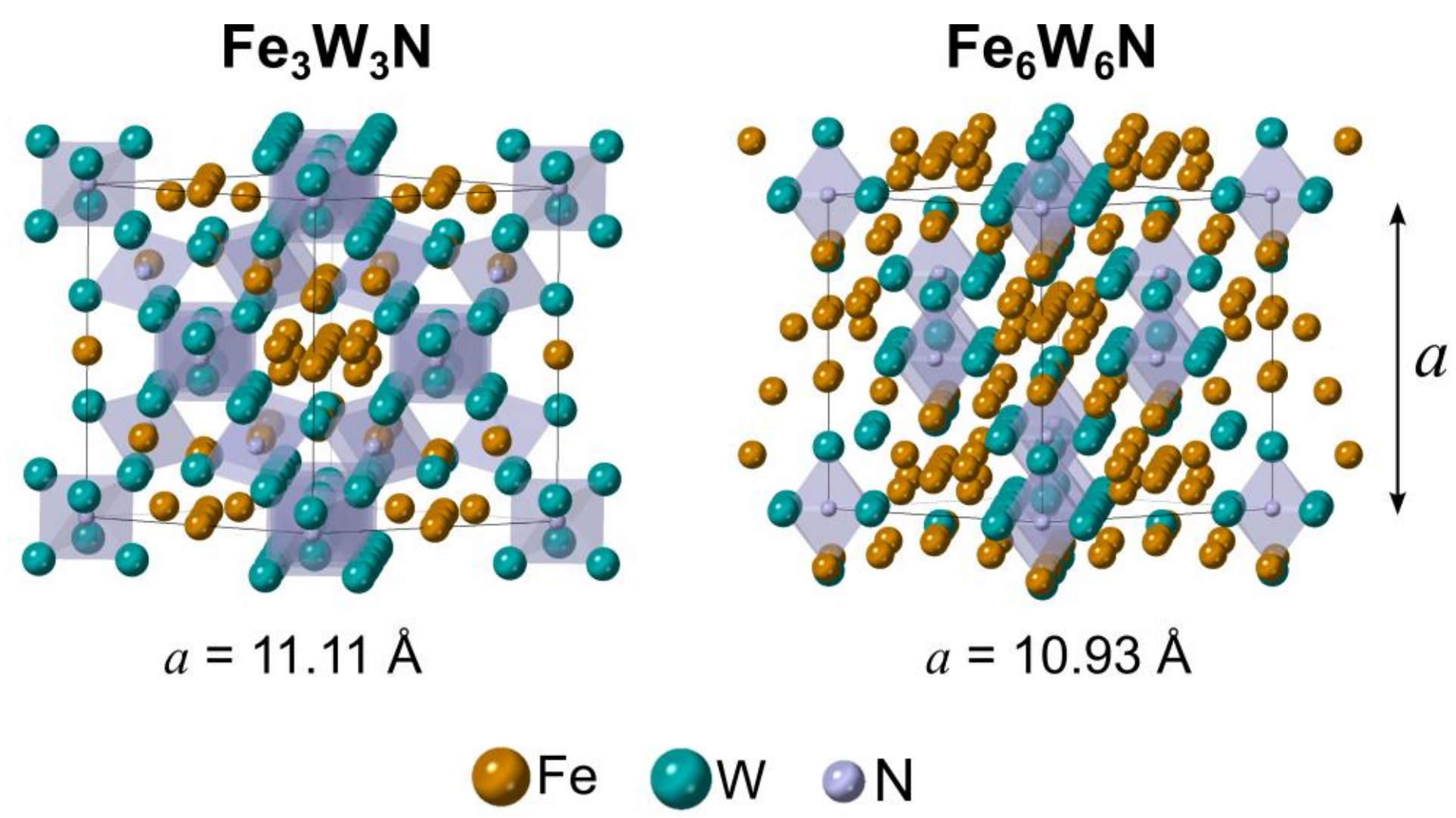

**Figure S8**. Crystal structure of $\eta$-nitride $Fe_3W_3N$ and $Fe_6W_6N$. The structures are based on ICSD entries 59255 and 208836, respectively.

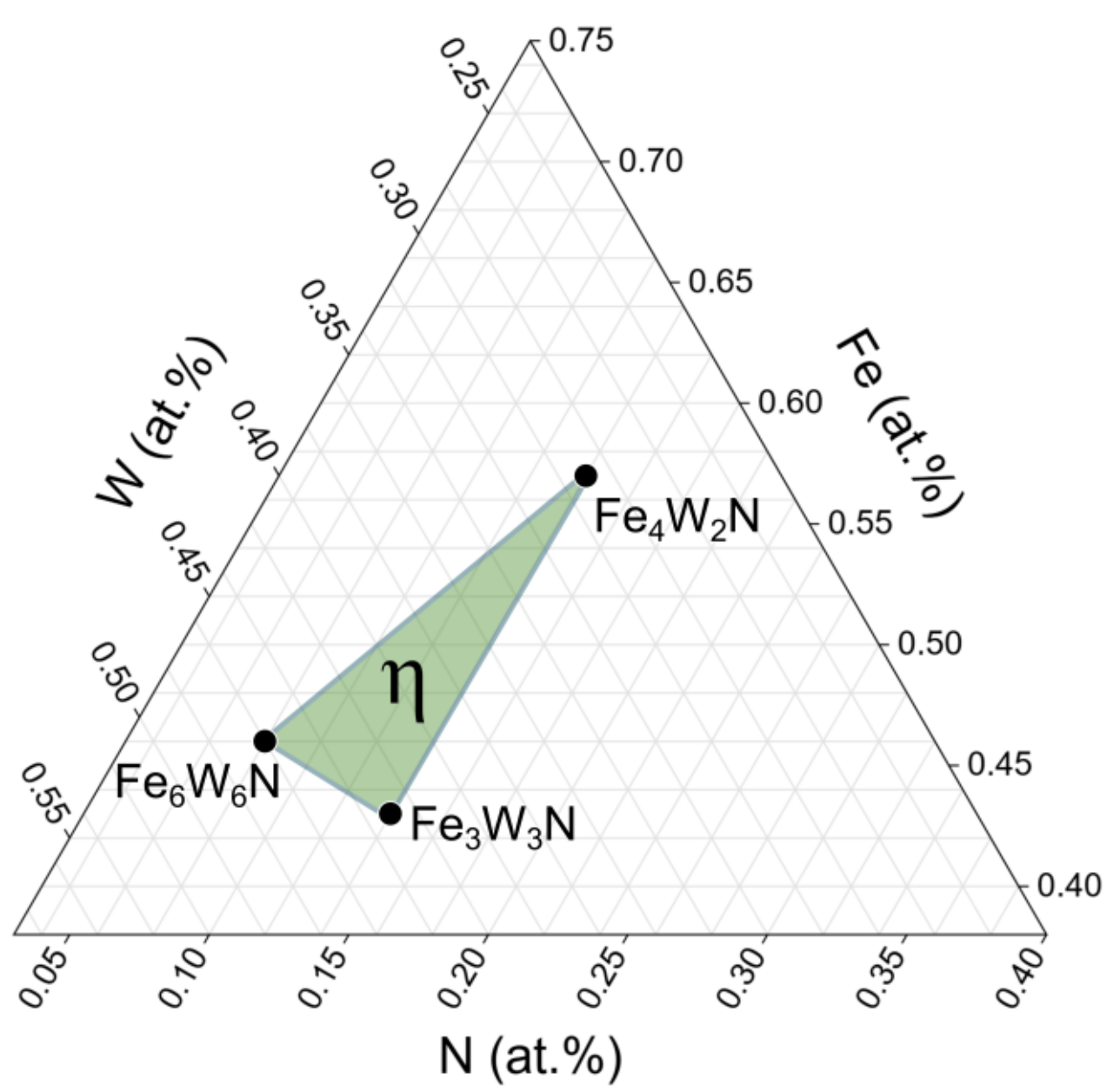

**Figure S9**. Ternary diagram Fe-W-N highlighting the tie line of compositions that exhibit the $\eta$ crystal structure.

## IV. Magnetic Measurements

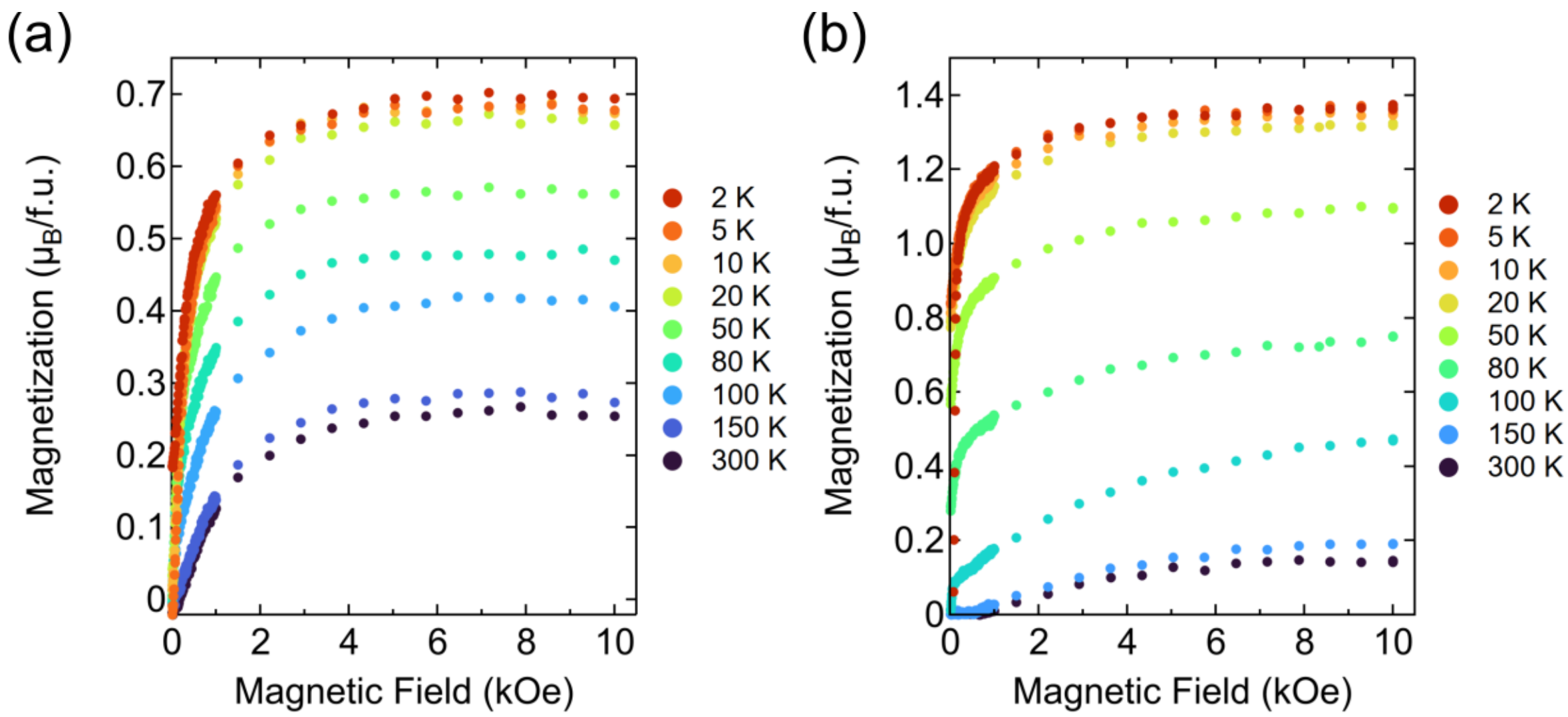

**Figure S10**. High-field magnetization measurements of Fe-rich films (a) $Fe_3W_3N$ (x = 0.64) and (b) $Fe_3Mo_3N$ (x = 0.59). In $Fe_3W_3N$, some additional slope correction at high field is applied due to some diamagnetic background bringing down the magnetization. In $Fe_3Mo_3N$, the magnetization does not seem to reach saturation at 10 kOe and keeps increasing linearly.

<u>**Additional details on methods for measuring and processing magnetic data of thin films**</u>

The magnetic properties of selected Fe-W-N and Fe-Mo-N thin films on Si/$SiN_x$ substrate were measured in a Physical a Quantum Design DynaCool Physical Property Measurement System (PPMS) using the AC measurement system (ACMS) option which is based on vibration sample magnetometry (VSM) technique. We note that a “sample” here include film + substrate. Each sample was diced into ~ 5 x 5 mm pieces. Before loading in the instrument, the actual sample dimensions were measured as well as the mass. The thickness of the film was estimated by XRF. The sample was then mounted to a quartz measurement rod using insulating varnish (model VGE-7031) on the backside of the sample. The varnish was left on air to cure for about 30 min, and the sample was then loaded in the instrument. To remove the contribution of the substrate and obtain the film properties only, a piece of bare Si/$SiN_x$ substrate around the same size was also measured in the same condition.

The magnetic moment of the film only $\mu_{film}$ is obtained by subtracting the magnetic moment of the substrate only $\mu_{substrate}$ (measured in units of emu) to the magnetic moment of the sample $\mu_{sample}$ (measured in units of emu). A scale factor, accounting for mass differences between the sample and the substrate pieces, is applied in the subtraction. The resulting moment (in units of emu) is expressed as:

$$\mu_{film} = \mu_{sample} - \frac{m_{sample}}{m_{substrate}} \mu_{substrate} \quad (1)$$

The next step is to normalize the magnetization of the film by its mass and ultimately by formula units of $Fe_3W_3N$ or $Fe_3Mo_3N$. Since it is not possible to measure the real mass of the film only, this conversion assumes that the film is fully composed of the material investigated (i.e. $Fe_3W_3N$ or $Fe_3Mo_3N$) and does not take into account any impurities or secondary phases.

First, the volume of the film $V_{film}$ is estimated using the area of the sample $A_{sample}$ and the thickness of the film $d_{film}$:

$$V_{film} = A_{sample} \times d_{film} \quad (2)$$

The mass of the film $m_{film}$ is then estimated from the theoretical density $\rho$ of the material investigated, assuming no other phases:

$$m_{film} = \rho \times V_{film} \quad (3)$$

The mass magnetization of the film $M_g$ (in emu/g) is obtained as:

$$M_g = \frac{\mu_{film}}{m_{film}} \quad (4)$$

Finally, the magnetization per formula unit is expressed in term of Bohr magneton ($\mu_B$/f.u.):

$$M = \frac{M_g}{9.274 \times 10^{-21}} \times \frac{M_{molar}}{N_A} \quad (5)$$

The second term on the right represents the mass of one formula unit of material where $M_{molar}$ is the molar mass of the material (in g/mol) and $N_A$ the Avogadro number (in $mol^{-1}$).

The magnetization can then be expressed per Fe atom by simply dividing per number of Fe per formula unit, i.e. 3 for $Fe_3W_3N$ or $Fe_3Mo_3N$.

In the case of magnetic susceptibility, the magnetic moment $\mu_{film}$ (in emu) is normalized by moles of formula unit of material and then divided by the applied field to obtain the susceptibility in unit of emu/mol-f.u.

## V. Mixed Chemical Potential – Composition Phase Diagrams

Thin film nitride synthesis of Fe-M-N ternary nitrides, where M = Mo/W, possess thermodynamic boundary conditions that are open to the exchange of nitrogen with an external reservoir, and closed to the metallic composition ratio of Fe/(Fe+M). Thus, phase stability is evaluated using a mixed nitrogen chemical potential – metallic composition phase diagram. Each phase $i$, in the ternary Fe-M-N system is defined by the all-intensive energy potential $\phi_i$ which is a concave hyper-plane in $\phi - \mu_{Fe} - \mu_M - \mu_N$ space with composition coefficients of $N_{Fe,i}$, $N_{M,i}$, and $N_{N,i}$. SciPy's half space intersection algorithm [1] is then used to compute the stability domain of each phase in $\mu_{Fe} - \mu_M - \mu_N$ space by computing the lower half-space envelop of all $\phi_i$ corresponding to all the phases.

Geometrically, the stability domains of single phases are 2-dimensional polygons in in $\mu_{Fe} - \mu_M - \mu_N$ space, with the two-phase coexistence regions being 1-dimensional edges where two polygons meet, and three-phase coexistence regions being vertices where three polygons meet. For the $\mu_N - x_{Fe/(Fe+M)}$ diagrams, the vertical single-phase lines are plotted by extracting the projection of the single-phase polygons onto the $\mu_N$ axis. The 2-phase coexistence rectangles are plotted by identifying pairs of polytopes with shared vertices that form 2-dimensional edges, with each edge defining a minimum and maximum $\mu_N$ which defines the height of the rectangle, while the two compositions of the coexisting phases define the rectangle width. Finally, the horizontal three phase coexistence lines are plotted by identifying triplets of polygons with a single shared vertex, which defines the $\mu_N$ value at which the horizontal line appears, while the three compositions of the coexisting phases define its horizontal extent.